\documentclass[journal,twoside,web]{ieeecolor}
\pdfoutput=1
\usepackage{generic}
\usepackage{graphicx}
\usepackage{amsmath}
\usepackage{amssymb}
\usepackage{nicefrac}
\usepackage{mathtools}
\usepackage{cite}
\usepackage{color}
\usepackage{booktabs}
\usepackage[flushleft]{threeparttable}
\usepackage{caption}
\makeatletter
\let\NAT@parse\undefined
\makeatother
\usepackage{hyperref}

\usepackage{fancyhdr}
\usepackage{xcolor}

\hypersetup{
	pdftitle={Schiller2023 - A Lyapunov function for robust stability of moving horizon estimation},
	pdfauthor={Julian D. Schiller, Simon Muntwiler, Johannes Köhler, Melanie N. Zeilinger, Matthias A. Müller},
	colorlinks,
	linkcolor={red!50!black},
	citecolor={blue!50!black},
	urlcolor={blue!80!black}
}

\def\doi{10.1109/TAC.2023.3280344}
\fancyfootoffset{-1em}
\fancypagestyle{copyright}{
	\cfoot{\vspace{-1.8em}\footnotesize\textcopyright{} 2023 IEEE. Personal use of this material is permitted.  Permission from IEEE must be obtained for all other uses, in any current or future media, including reprinting/republishing this material for advertising or promotional purposes, creating new collective works, for resale or redistribution to servers or lists, or reuse of any copyrighted component of this work in other works.}
	\chead{\footnotesize This article has been accepted for publication in IEEE Transactions on Automatic Control. This is the author's version which has not been fully edited and content may change prior to final publication. Citation information: \href{http://dx.doi.org/\doi}{DOI \doi}}
}

\newtheorem{assumption}{Assumption}
\newtheorem{definition}{Definition}
\newtheorem{theorem}{Theorem}
\newtheorem{proposition}{Proposition}

\newtheorem{corollary}{Corollary}
\newtheorem{remark}{Remark}
\newtheorem{claim}{Claim}

\title{A Lyapunov function for robust stability of moving horizon estimation
\thanks{$^{\star}$Julian D. Schiller and Simon Muntwiler contributed equally to this paper.}
\thanks{This work was funded by the Deutsche Forschungsgemeinschaft (DFG, German Research Foundation) --  426459964, and by the Bosch Research Foundation im Stifterverband. \textit{(Corresponding authors: Julian D. Schiller; Simon Muntwiler)}}
\thanks{Julian D. Schiller and Matthias A. M\"uller are with the Institute of Automatic Control, Leibniz University Hannover, 30167
	Hannover, Germany. (e-mail: $\{$schiller,mueller$\}$@irt.uni-hannover.de).}
\thanks{Simon Muntwiler, Johannes K\"ohler, and Melanie N. Zeilinger are with the Institute for Dynamic Systems and Control, ETH Z\"urich, 8092 Z\"urich, Switzerland.
	(e-mail: $\{$simonmu,jkoehle,mzeilinger$\}$@ethz.ch).}}
\author{\parbox{\linewidth}{\centering Julian D. Schiller$^{\star}$, Simon Muntwiler$^{\star}$, Johannes Köhler, Melanie N. Zeilinger, Matthias A. Müller}}
\begin{document}
\maketitle
\thispagestyle{copyright}
\pagestyle{empty}


\begin{abstract}
We provide a novel robust stability analysis for moving horizon estimation (MHE) using a Lyapunov function.
Additionally, we introduce linear matrix inequalities (LMIs) to verify the necessary incremental input/output-to-state stability (\mbox{$\boldsymbol{\delta}$-IOSS}) detectability condition.
We consider an MHE formulation with time-discounted quadratic objective for nonlinear systems admitting an exponential $\boldsymbol{\delta}$-IOSS Lyapunov function.
We show that with a suitable parameterization of the MHE objective, the $\boldsymbol{\delta}$-IOSS Lyapunov function serves as an $\boldsymbol{M}$-step Lyapunov function for MHE.
Provided that the estimation horizon is chosen large enough, this directly implies exponential stability of MHE.
The stability analysis is also applicable to full information estimation, where the restriction to \emph{exponential} $\boldsymbol{\delta}$-IOSS can be relaxed.
Moreover, we provide simple LMI conditions to systematically derive $\boldsymbol{\delta}$-IOSS Lyapunov functions, which allows us to easily verify $\boldsymbol{\delta}$-IOSS for a large class of nonlinear detectable systems.
This is useful in the context of MHE in general, since most of the existing nonlinear (robust) stability results for MHE depend on the system being $\boldsymbol{\delta}$-IOSS (detectable).
In combination, we thus provide a framework for designing MHE schemes with guaranteed robust exponential stability.
The applicability of the proposed methods is demonstrated with a nonlinear chemical reactor process and a 12-state quadrotor model.
\end{abstract}
\begin{IEEEkeywords}
Moving horizon estimation; State estimation; Incremental system properties
\end{IEEEkeywords}
\vspace{1ex}
\section{Introduction}

State estimation for nonlinear systems based on noisy output measurements is a challenging problem of high practical relevance. 
The design of corresponding state observers is an active field of research, with recent results using differential dynamics and reduced coordinates~\cite{manchester2021observer,sanfelice2021metric}, and observers for constrained nonlinear systems with a quadratic Lyapunov function~\cite{astolfi2022constrained}.
An optimization-based approach to nonlinear state estimation is moving horizon estimation (MHE)~\cite{allan2019moving}, \cite[Chap. 4]{rawlings2020model}.
Our main contribution is twofold: We provide a robust stability analysis for MHE using a Lyapunov function, and we provide simple linear matrix inequality (LMI) conditions to verify the underlying incremental input/output-to-state stability ($\delta$-IOSS) detectability assumption.

\subsubsection*{Related work}
Based on an observability assumption, nominal stability of MHE without prior weighting and with a constant quadratic form as prior weighting were shown in~\cite{michalska1995moving} and~\cite{alessandri2008moving}, respectively.
In~\cite{rao2003constrained}, an approximation of the arrival cost was used as prior weighting to mimic the effect of the neglected past measurements.
Alternative approaches to stability of MHE are based on the inclusion of a robustly stabilizing observer~\cite{gharbi2021proximity,schiller2021suboptimal}.

More general robust stability results of MHE are based on $\delta$-IOSS as a notion of detectability~\cite{sontag1997output}.
In particular, in~\cite{rawlings2012optimization}, this detectability assumption was used to show suitable stability properties for full information estimation (FIE) in case of convergent disturbances.
This restriction of convergent disturbances has been relaxed in~\cite{muller2016nonlinear,hu2017robust} for MHE, by introducing an additional \textit{max}-term penalizing the largest stage cost into the MHE objective.
In~\cite{allan2019moving,muller2017nonlinear}, robust stability was shown without the additional \textit{max}-term, allowing for standard quadratic objective functions.
However, in both~\cite{allan2019moving} and \cite{muller2017nonlinear}, the resulting estimation error bounds become worse for larger MHE horizons and the derived stability properties do not hold globally (cf. \cite[Rem. 2]{allan2019moving}).
A first step toward a generalized stability analysis of MHE based on Lyapunov theory was presented in~\cite{allan2019lyapunov,allan2021FIE} by introducing a Lyapunov-like function for the stability analysis of FIE, using the fact that robust \textit{exponential} stability of FIE also implies stability of MHE for a sufficiently large horizon, cf. \cite{allan2021FIE} and compare also \cite{hu2017robust,knuefer2021MHE,hu2021MHE}.
Finally, the use of $\delta$-IOSS and further a time-discounted objective function allowed to show robust stability of FIE and MHE~\cite{knufer2018robust,knuefer2021MHE}.

\subsubsection*{Contribution}
In this paper, we present an MHE scheme with exponential discounting, but otherwise standard quadratic stage cost and prior weighting, for systems admitting an \textit{exponential} $\delta$-IOSS Lyapunov function (Section~\ref{sec:mhe_formulation}).
Provided the time-discounting factor in the MHE objective satisfies a certain condition based on the $\delta$-IOSS Lyapunov function, any quadratic objective can be considered in the MHE problem.
For the stated MHE scheme, we present a robust stability analysis based on a Lyapunov function for MHE (Section~\ref{sec:MHE_stability}), resulting in theoretical guarantees which improve as the MHE horizon increases.
In particular, we show that the current $\delta$-IOSS Lyapunov function is bounded by a past $\delta$-IOSS Lyapunov function, the current value function, and a bound depending on the disturbances during the MHE horizon window.
This result is partially motivated by a similar bound derived in~\cite{koehler2021Output} in the context of output-feedback model predictive control (MPC).
Based on the above bound, we show that the $\delta$-IOSS Lyapunov function is an $M$-step Lyapunov function for the MHE scheme (Theorem~\ref{thm:MHE_Lyapunov_function}), where $M$ is the horizon length of the MHE problem. 
Thereby, a sufficient lower bound on the horizon length $M$ is obtained, which ensures a decrease in the Lyapunov function over $M$ time steps and thus directly ensures robust stability (Corollary~\ref{cor:MHE_RGES}).
The proposed analysis is directly applicable to FIE, where we can additionally relax the restriction to \textit{asymptotic} (instead of exponential) $\delta$-IOSS (Section~\ref{sec:FIE}).
In Section~\ref{sec:discussion}, we provide a detailed discussion of our proposed stability analysis with respect to recent MHE and FIE stability results~\cite{allan2019moving,hu2017robust,muller2017nonlinear,allan2019lyapunov,allan2021FIE,allan2020lyapunov,knufer2018robust,knuefer2021MHE,hu2021MHE}.
In particular, compared to existing Lyapunov-like techniques \cite{allan2019lyapunov,allan2021FIE}, our framework allows for a much simpler Lyapunov function and corresponding robust stability analysis.
Moreover, we show that the proposed condition on the horizon length for guaranteed robust stability of MHE is (significantly) less conservative than the corresponding conditions required by recent robust stability results from the literature on nonlinear MHE, in particular, \cite{allan2019moving,knuefer2021MHE,allan2021FIE}, compare Table~\ref{tab:horizons}.

As second contribution, we provide a systematic approach to verify exponential $\delta$-IOSS for a large class of nonlinear detectable systems based on their differential dynamics (Section~\ref{sec:dIOSS}).
More precisely, we provide sufficient conditions for a quadratically bounded $\delta$-IOSS Lyapunov function that can be easily cast in terms of LMIs (Theorem~\ref{thm:dIOSS_Lyap}).
This contribution is of particular relevance for all MHE schemes that rely on $\delta$-IOSS, e.g.,  \cite{allan2021detect,knuefer2020time,knufer2018robust,knuefer2021MHE,muller2017nonlinear,allan2019lyapunov,hu2021MHE,allan2021FIE}, but also for the verification of more general system properties such as incremental dissipativity.

Overall, the proposed $\delta$-IOSS verification, and in particular the resulting $\delta$-IOSS Lyapunov function, allows us to choose a beneficial quadratic prior weighting such that the MHE scheme from Section~\ref{sec:mhe_formulation} is robustly exponentially stable with relatively short horizon.
The applicability of the proposed framework is demonstrated in Section~\ref{sec:num} with a nonlinear chemical reactor process from~\cite{rawlings2020model,tenny2002efficient}, and a \mbox{12-state} quadrotor model.
For these examples, we can rigorously show satisfaction of the posed conditions and derive a short (practical) horizon bound that guarantees robust exponential stability of MHE.

\subsubsection*{Notation}
Let the non-negative real numbers be denoted by $\mathbb{R}_{\geq 0}$, the set of integers by $\mathbb{I}$, the set of all integers greater than or equal to $a$ for some $a \in \mathbb{R}$ by $\mathbb{I}_{\geq a}$, and the set of integers in the interval $[a,b]$ for some $a,b\in\mathbb{R}$ with $a\le b$ by $\mathbb{I}_{[a,b]}$.
Let $\|x\|$ denote the Euclidean norm of the vector $x \in \mathbb{R}^n$.
The quadratic norm with respect to a positive definite matrix $Q=Q^\top$ is denoted by $\|x\|_Q^2=x^\top Q x$, and the minimal and maximal eigenvalues of $Q$ are denoted by $\lambda_{\min}(Q)$ and $\lambda_{\max}(Q)$, respectively.
The maximum generalized eigenvalue of positive definite matrices $A=A^\top$ and $B=B^\top$ is denoted as $\lambda_{\max}(A,B)$, i.e., the largest scalar $\lambda$ satisfying $\det (A-\lambda B) = 0$.
The identity matrix is denoted by $I_n\in\mathbb{R}^{n\times n}$. 
A function $\alpha:\mathbb{R}_{\geq 0}\rightarrow\mathbb{R}_{\geq 0}$ is of class $\mathcal{K}$ if it is continuous, strictly increasing, and satisfies $\alpha(0)=0$. 
If $\alpha$ is additionally unbounded, it is of class $\mathcal{K}_{\infty}$.
We denote the class of functions $\theta:\mathbb{I}_{\geq 0}\rightarrow \mathbb{R}_{\geq 0}$ that are continuous, non-increasing, and satisfy $\lim_{t\rightarrow\infty}\theta(t)=0$ by $\mathcal{L}$.
By $\mathcal{KL}$, we denote the functions $\beta:\mathbb{R}_{\geq 0}\times\mathbb{I}_{\geq 0}\rightarrow\mathbb{R}_{\geq 0}$ with $\beta(\cdot,t)\in\mathcal{K}$ and $\beta(r,\cdot)\in\mathcal{L}$ for any fixed $t\in\mathbb{I}_{\geq 0}$, $r\in\mathbb{R}_{\geq 0}$.
\section{Problem setup and preliminaries}
 \label{sec:setup}
 
We consider the discrete-time, nonlinear perturbed system
\begin{subequations}
	\label{eq:sys_w}
	\begin{align}
	\label{eq:sys_w_1}
	x_{t+1}&=f(x_t,u_t,w_t),\\
	\label{eq:sys_w_2}
	y_t&=h(x_t,u_t,w_t), 
	\end{align}
\end{subequations}
with state $x_t\in\mathbb{R}^n$, control input $u_t\in\mathbb{R}^m$, disturbance $w_t\in\mathbb{R}^{q}$, noisy output measurement $y_t\in\mathbb{R}^p$, and time $t\in\mathbb{I}_{\geq 0}$.
The nonlinear continuous functions  $f:\mathbb{R}^n\times\mathbb{R}^m\times\mathbb{R}^q\rightarrow\mathbb{R}^n$ and  $h:\mathbb{R}^n\times\mathbb{R}^m\times\mathbb{R}^q\rightarrow\mathbb{R}^p$ represent the system dynamics and the output equation, respectively.
Note that the generalized disturbance $w$ accounts for both the process disturbance (\ref{eq:sys_w_1}) and the measurement noise (\ref{eq:sys_w_2}) to keep the presentation concise.
This general formulation also covers the standard setting of independent process disturbance and measurement noise as special case, cf. the numerical examples in Section~\ref{sec:num} below.
Since we only consider the estimation problem, the control input $u$ is treated as a known external variable. 

Given some initial guess $\hat{x}_0$ of the true state $x_0$, the main objective is to obtain, at each time $t \in \mathbb{I}_{\geq0}$, an estimate $\hat{x}_t$ of the current state $x_t$.
We consider the general case where we may know some additional information of the form\footnote{To simplify the notation, we restrict our analysis to decoupled constraint sets here. The stability properties for MHE and FIE presented in Section~\ref{sec:estimation} remain valid also for coupled constraints, i.e., $(x_t,u_t,w_t,y_t)\in\mathbb{Z}$.
Note that in the unconstrained case, i.e., when no additional information of the form~\eqref{eq:constraints} is available, the results presented in sections~\ref{sec:estimation} and~\ref{sec:dIOSS} below remain valid with $\mathbb{X}=\mathbb{R}^n$, $\mathbb{U}=\mathbb{R}^m$, $\mathbb{W}=\mathbb{R}^q$, and $\mathbb{Y}=\mathbb{R}^p$.}
\begin{equation}\label{eq:constraints}
	(x_t,u_t,w_t,y_t)\in\mathbb{X}\times\mathbb{U}\times\mathbb{W}\times\mathbb{Y}\eqqcolon\mathbb{Z}, \quad t\in\mathbb{I}_{\geq 0},
\end{equation}
with sets $\mathbb{X}\subseteq\mathbb{R}^n$, $\mathbb{U}\subseteq\mathbb{R}^m$, $\mathbb{W}\subseteq\mathbb{R}^q$, and $\mathbb{Y}\subseteq\mathbb{R}^p$.
Note that $\mathbb{Z}$ does not represent a set of constraints in the sense of a control problem, but rather the domain of real system trajectories.
This typically arises from physical nature, e.g., mechanically imposed limits on joint angles or a measurement device; or non-negativity of the absolute temperature, partial pressures, or concentrations of species in a chemical reaction.
Taking such information into account can often significantly improve the estimation results, cf. \cite[Sec. 4.4]{rawlings2020model}. 

In order to establish robust stability of the proposed MHE scheme, an appropriate detectability assumption is required.
To this end, we consider the following $\delta$-IOSS Lyapunov function.
\begin{definition}[$\delta$-IOSS Lyapunov function~\protect{\cite[Def.~2.9]{allan2021detect}}]
	\label{def:IOSS_Lyap}
	A function $W_\delta:\mathbb{R}^n\times\mathbb{R}^n\rightarrow\mathbb{R}_{\geq 0}$ is a $\delta$-IOSS Lyapunov function if there exist $\alpha_1,\alpha_2\in\mathcal{K}_{\infty}$, $\sigma_{w},\sigma_{y}\in\mathcal{K}$, and $\eta\in[0,1)$ such that
	\begin{subequations}
		\label{eq:IOSS_Lyap}
		\begin{align}
		\label{eq:IOSS_Lyap_1}
		&\alpha_1(\|x-\tilde{x}\|)\leq W_\delta(x,\tilde{x}) \leq \alpha_2(\|x-\tilde{x}\|),\\	
		&W_\delta(f(x,u,w),f(\tilde{x},u,\tilde{w}))  \nonumber \\ &\quad \leq \eta W_\delta(x,\tilde{x}) +\sigma_{w}(\|w-\tilde{w}\|)+\sigma_{y}(\|y-\tilde{y}\|), \label{eq:IOSS_Lyap_2}
		\end{align}
	\end{subequations}
	for all $(x,u,w,y),(\tilde{x},u,\tilde{w},\tilde{y})\in\mathbb{Z}$, where $y=h(x,u,w)$ and $\tilde{y}=h(\tilde{x},u,\tilde{w})$.
\end{definition}

Definition~\ref{def:IOSS_Lyap} is equivalent to a corresponding $\delta$-IOSS property involving general $\mathcal{KL}$-functions (cf. \cite[Thm.~3.2]{allan2021detect}\footnote{%
	We conjecture that the converse Lyapunov results from \cite[Thm.~3.2]{allan2021detect}, where $y=h(x)$ is assumed, remain valid for our more general nonlinear setup (\ref{eq:sys_w}).
}), which became standard as a description of nonlinear detectability in the context of MHE in recent years \cite{allan2021detect,knuefer2020time,knufer2018robust,knuefer2021MHE,muller2017nonlinear,allan2019lyapunov,hu2021MHE,allan2021FIE}. 
Note that $\delta$-IOSS is both necessary (cf.~\cite[Prop.~3]{knuefer2020time}, \cite[Prop.~2.6]{allan2021detect}) and sufficient (cf.~\cite[Thm.~11]{knuefer2021MHE}) for the existence of robustly stable state estimators as characterized below by Definition~\ref{def:RGAS}.

In order to provide stability guarantees with a finite-horizon, we assume quadratic bounds and supply rates, which implies an \textit{exponential} $\delta$-IOSS condition, compare also the discussion in Section~\ref{sec:discussion} below.
\begin{assumption}[Exponential $\delta$-IOSS]\label{ass:IOSS_quadratically_bounded}
	The system \eqref{eq:sys_w} admits a $\delta$-IOSS Lyapunov function $W_\delta$ according to Definition~\ref{def:IOSS_Lyap} with quadratic bounds and supply rates, i.e., there exist $P_1,P_2\succ0$ and $Q,R\succeq0$ such that
	\begin{subequations}
		\label{eq:dIOSS_exp}
		\begin{align}\label{eq:dIOSS_quad_bounds}
			&\|x-\tilde{x}\|_{P_1}^2 \leq W_\delta(x,\tilde{x}) \leq \|x-\tilde{x}\|_{P_2}^2, \\
			&\ W_\delta(f(x,u,w),f(\tilde{x},u,\tilde{w})) \nonumber \\ 
			&\quad \leq \eta W_\delta(x,\tilde{x})
			 +\|w-\tilde{w}\|_Q^2 +\|y-\tilde{y}\|_R^2 \label{eq:dIOSS_quad_gains}
		\end{align}
	\end{subequations}
	for all $(x,u,w,y),(\tilde{x},u,\tilde{w},\tilde{y})\in\mathbb{Z}$, where $y=h(x,u,w)$ and $\tilde{y}=h(\tilde{x},u,\tilde{w})$.
\end{assumption}

The explicit computation of such a quadratically bounded $\delta$-IOSS Lyapunov function is discussed in detail in Section~\ref{sec:dIOSS}.
In the following, we utilize $\delta$-IOSS to design an MHE framework that is robustly stable by means of the following notion.

\begin{definition}[\protect{RGAS~\cite[Def.~2.3]{allan2021detect}, RGES \cite[Def. 1]{knufer2018robust}}]
	\label{def:RGAS}
	A state estimator for system~\eqref{eq:sys_w} is robustly globally asymptotically stable (RGAS) if there exist $\beta_{1},\beta_{2}\in\mathcal{KL}$ such that the resulting state estimates $\hat{x}_t$ satisfy
	\begin{align}	
	\label{eq:RGAS}	
	\|x_{t}-\hat{x}_{t}\| &\leq \max \lbrace\beta_{1}(\|x_0-\hat{x}_0\|,t),\\	
	&\ \ \ \max_{j\in\mathbb{I}_{[0,t-1]}}\beta_{2}(\|w_j\|,t-j-1)\rbrace\nonumber
	\end{align}
	for all $t\in\mathbb{I}_{\geq0}$, all initial conditions $x_0,\hat{x}_0\in\mathbb{X}$, and every trajectory $(x_t,u_t,w_t,y_t)_{t=0}^\infty$ satisfying~\eqref{eq:sys_w}-\eqref{eq:constraints}.
	If additionally $\beta_{1}(r,t)=C_1\lambda_1^tr$ and $\beta_{2}(r,t)=C_2\lambda_2^tr$  with $\lambda_1,\lambda_2\in [0,1)$ and $C_1,C_2>0$, then the state estimator is robustly globally exponentially stable (RGES).
\end{definition}

This definition of robust stability is often used in the recent MHE literature and has already been adequately studied in, e.g., \cite{rawlings2020model,knuefer2021MHE,allan2020lyapunov,knuefer2020time,allan2021detect}.
Note that this characterization is particularly suitable for MHE and FIE, since it directly implies that the estimation error converges to zero if the disturbances vanish \cite[Prop. 4.3]{rawlings2020model}, which would not immediately be the case using robust stability notions without time-discounting, compare \cite{muller2017nonlinear,allan2019moving,hu2017robust}.

\section{Lyapunov function for moving horizon estimation}
\label{sec:estimation}

In this section, we start by introducing the proposed MHE formulation with a cost function related to the $\delta$-IOSS Lyapunov function in Section~\ref{sec:mhe_formulation}.
We then present the stability analysis of the MHE scheme by showing that it admits a Lyapunov function in Section~\ref{sec:MHE_stability}.
In particular, we establish a relation between the $\delta$-IOSS Lyapunov function and the value function of the MHE problem.
Based on this relation, we show that the $\delta$-IOSS Lyapunov function serves as an $M$-step Lyapunov function for MHE, where $M$ is the horizon length of the MHE.
In Section~\ref{sec:FIE}, an extension of the analysis to FIE is presented, which allows us to avoid Assumption~\ref{ass:IOSS_quadratically_bounded}, i.e., \emph{exponential} $\delta$-IOSS.
Finally, a detailed discussion of the differences between the presented stability analysis with related approaches is presented in Section~\ref{sec:discussion}.

\subsection{Moving Horizon Estimator Formulation}\label{sec:mhe_formulation}
At time $t$, the MHE scheme considers past input and output data $(u,y)$  in a window of length $M_t=\min\{t,M\}$, with $M\in\mathbb{I}_{\geq 0}$, and the past estimate\footnote{This choice is typically called filtering prior, cf.\cite{allan2019moving}, \cite[Chap. 4]{rawlings2020model}.} $\hat{x}_{t-M_t}$.
Thereby, the MHE optimizes over the initial estimate $\hat{x}_{t-M_t|t}$ and a sequence of $M_t$ disturbance estimates $\hat{w}_{\cdot|t}=\left\{\hat{w}_{j|t}\right\}_{j=t-M_t}^{t-1}$.
Combined, the initial estimate and sequence of disturbance estimates define a sequence of state estimates $\hat{x}_{\cdot|t}=\left\{\hat{x}_{j|t}\right\}_{j=t-M_t}^{t}$ and a sequence of output estimates $\hat{y}_{\cdot|t}=\left\{\hat{y}_{j|t}\right\}_{j=t-M_t}^{t-1}$.
The objective of this optimization-based state estimation problem is to minimize the following cost function
\begin{align}
		&V_{\mathrm{MHE}}(\hat{x}_{t-M_t|t},\hat{w}_{\cdot|t},\hat{y}_{\cdot|t},t) = 2\eta^{M_t}\|\hat{x}_{t-M_t|t}-\hat{x}_{t-M_t}\|_{P_2}^2 \nonumber \\
		& \quad +\sum_{j=1}^{M_t}\eta^{j-1}\left(2\|\hat{w}_{t-j|t}\|_Q^2+\|\hat{y}_{t-j|t}-y_{t-j}\|_R^2 \right), \label{eq:MHE_objective}
\end{align}
where $\eta$, $Q$, $R$, and $P_2$ are based on the exponential $\delta$-IOSS property and corresponding Lyapunov function $W_\delta$ according to Assumption~\ref{ass:IOSS_quadratically_bounded}.
In fact, provided the discount factor $\eta$ in the MHE cost~\eqref{eq:MHE_objective} is chosen such that $1-\eta>0$ is sufficiently small, the theoretical analysis in Section~\ref{sec:MHE_stability} below remains valid for any positive definite matrices $Q$, $R$, and $P_2$ in the MHE cost~\eqref{eq:MHE_objective}, compare also Remark~\ref{rem:cost} below.
A similar time-discounted MHE cost has been previously suggested in~\cite{knufer2018robust}, compare also~\cite{knuefer2021MHE} for a more general asymptotic discounting.
Except for the discount factor~$\eta$, the cost~\eqref{eq:MHE_objective} allows for standard (quadratic) MHE stage costs.
The use of a discount factor in the cost~\eqref{eq:MHE_objective} allows us to obtain tighter upper bounds in~\eqref{eq:RGAS}, compared to, e.g.,~\cite{allan2019moving,muller2017nonlinear} where the bounds deteriorate with a larger horizon length $M$ (compare also the discussion in Section~\ref{sec:discussion} below).
The state estimate at time step $t$ is then obtained by solving the following nonlinear program (NLP)
\begin{subequations}\label{eq:MHE_IOSS}
\begin{align}\label{eq:MHE_IOSS_cost}
&\min_{\hat{x}_{t-M_t|t},\hat{w}_{\cdot|t}} V_{\mathrm{MHE}}(\hat{x}_{t-M_t|t},\hat{w}_{\cdot|t},\hat{y}_{\cdot|t},t) \\ \label{eq:MHE_IOSS_1}
&\hspace{0.75cm} \text{s.t. }\hat{x}_{j+1|t}=f(\hat{x}_{j|t},u_{j},\hat{w}_{j|t}),j\in\mathbb{I}_{[t-M_t,t-1]},\\
\label{eq:MHE_IOSS_2}
&\hspace{1.3cm} \hat{y}_{j|t}=h(\hat{x}_{j|t},u_{j},\hat{w}_{j|t}),j\in\mathbb{I}_{[t-M_t,t-1]},\\
\label{eq:MHE_IOSS_3}
&\hspace{1.3cm}\hat{w}_{j|t} \in\mathbb{W},\hat{y}_{j|t}\in\mathbb{Y}, j\in\mathbb{I}_{[t-M_t,t-1]}, \\
\label{eq:MHE_IOSS_4}
&\hspace{1.3cm}\hat{x}_{j|t}\in\mathbb{X}, j\in\mathbb{I}_{[t-M_t,t]}.
\end{align}
\end{subequations}
We denote a minimizer\footnote{
	A (local) minimizer to~\eqref{eq:MHE_IOSS} exists under mild assumptions, e.g., if the cost function~\eqref{eq:MHE_objective} is assumed to be positive definite, i.e., $Q,R,P_2\succ0$, or the sets $\mathbb{X,W}$ to be compact, cf.~\cite[App. A.11]{rawlings2020model}.
} to~\eqref{eq:MHE_IOSS} by $\hat{x}^*_{t-M_t|t}$, $\hat{w}^*_{\cdot|t}$, and the corresponding estimated state and output trajectories as $\hat{x}^*_{\cdot|t}$, $\hat{y}^*_{\cdot|t}$, respectively.
The resulting state estimate at time step $t$ is given by
\begin{equation}\label{eq:MHE_estimate}
	\hat{x}_t = \hat{x}_{t|t}^*,
\end{equation}
and the estimation error at time step $t$, i.e., the difference between the true system state and the state estimate~\eqref{eq:MHE_estimate}, as
\begin{equation}\label{eq:MHE_estimate_error}
	\hat{e}_t = x_t - \hat{x}_t.
\end{equation}
The MHE estimator~\eqref{eq:MHE_IOSS} is then applied in a receding horizon fashion, i.e., at each time step $t$, the current state estimate~\eqref{eq:MHE_estimate} is obtained by solving the MHE problem~\eqref{eq:MHE_IOSS} based on the $M_t$ most recent output measurements.

\begin{remark}[Parameterization of MHE objective]\label{rem:cost}
	Assumption~\ref{ass:IOSS_quadratically_bounded} is invariant with respect to scaling of the $\delta$-IOSS Lyapunov function $W_\delta$.
	Specifically, if there exists a $\delta$-IOSS Lyapunov function with decay rate $\eta$ satisfying Assumption~\ref{ass:IOSS_quadratically_bounded}, the assumption can be satisfied for any $\tilde{\eta}\ge \eta$, with $\tilde{\eta}<1$, and any positive definite matrices $P_2$, $R$, and $Q$.
	Consequently, the cost of the MHE~\eqref{eq:MHE_objective} can be parameterized with any positive definite matrices $P_2$, $R$, and $Q$, and any discount factor $\tilde{\eta}$ provided that $1-\tilde{\eta}>0$ is chosen sufficiently small (see also \cite{knuefer2021MHE,schiller2021suboptimal} for a similar discussion).
	However, the choice of the matrix $P_2$ influences the minimal MHE horizon length required for RGES and the resulting error bounds (compare Section~\ref{sec:MHE_stability} and Theorem~\ref{thm:MHE_Lyapunov_function} below).
	If the ratio between $P_1$ and $P_2$ improves, i.e., the largest generalized eigenvalue $\lambda_{\max}(P_2,P_1)$ of $P_2$ and $P_1$ approaches 1, the horizon $M$ can be chosen smaller and the resulting estimation error bounds are less conservative.
	Consequently, choosing the prior weighting similar to the $\delta$-IOSS Lyapunov function improves the ratio between $P_1$ and $P_2$.
	Similar considerations also apply to Definition~\ref{def:IOSS_Lyap} and the FIE cost function~\eqref{eq:FIE_objective} used in Section~\ref{sec:FIE} below, where a scaling of the functions $\alpha_2,\sigma_{w},\sigma_{y}$ can be considered.
\end{remark}

\subsection{Lyapunov-based Stability Analysis}\label{sec:MHE_stability}
In the following, we elaborate how the specific choice of the cost function~\eqref{eq:MHE_objective} based on the $\delta$-IOSS Lyapunov function~$W_\delta$ results in $W_{\delta}$ being an $M$-step Lyapunov function for MHE.
Given the horizon $M$ satisfies some lower bound, it follows directly that the proposed MHE scheme is an RGES state estimator according to Definition~\ref{def:RGAS}, compare Theorem~\ref{thm:MHE_Lyapunov_function} below.
We start by showing the relation of $W_{\delta}(\hat{x}_t,x_t)$ and the value function $V_{\mathrm{MHE}}(\hat{x}_{t-M_t|t}^*,\hat{w}_{\cdot|t}^*,\hat{y}_{\cdot|t}^*,t)$ of~\eqref{eq:MHE_IOSS} in the following Proposition.

\begin{proposition}
	\label{prop:estimation_error_bound}
	Let Assumption~\ref{ass:IOSS_quadratically_bounded} hold.
	Then, for all $t\in\mathbb{I}_{\ge 0}$, the state estimate~$\hat{x}_t$~\eqref{eq:MHE_estimate} satisfies
	\begin{align}\label{eq:W_delta_value_function}
		W_{\delta}(&\hat{x}_t,x_t) \le 2\eta^{M_t} \lambda_{\max}(P_2,P_1) W_\delta(\hat{x}_{t-M_t},x_{t-M_t}) \\
		&+ V_{\mathrm{MHE}}(\hat{x}_{t-M_t|t}^*,\hat{w}_{\cdot|t}^*,\hat{y}_{\cdot|t}^*,t)  +2\sum_{j=1}^{M_t}\eta^{j-1} \|w_{t-j}\|_Q^2. \nonumber
	\end{align}
\end{proposition}
\begin{proof}
	Due to the constraints~\eqref{eq:MHE_IOSS_1}-\eqref{eq:MHE_IOSS_4} in the MHE optimization problem~\eqref{eq:MHE_IOSS}, we have at each time step $t$ that  $(\hat{x}^*_{j|t},u_{j},\hat{w}_{j|t}^*,\hat{y}_{j|t}^*)\in\mathbb{Z}$ for all $j\in\mathbb{I}_{[t-M_t,t-1]}$, $\hat{x}_{t|t}^*\in\mathbb{X}$, and the estimated trajectories satisfy~(\ref{eq:sys_w}).
	Thus, we can apply Inequality~\eqref{eq:dIOSS_quad_gains} $M_t$ times, which together with application of the upper bound~\eqref{eq:dIOSS_quad_bounds} leads to
	\begin{align}
	&\, W_\delta(\hat{x}_t,x_t) \nonumber \\ 
	\leq &\ \sum_{j=1}^{M_t}\eta^{j-1}\left(2\|\hat{w}^*_{t-j|t}\|_Q^2+2\|w_{t-j}\|_Q^2 +\|\hat{y}_{t-j|t}^*-y_{t-j}\|_R^2 \right) \nonumber \\
	&+ \eta^{M_t} \|\hat{x}_{t-M_t|t}^*-x_{t-M_t} \|_{P_2}^2, \label{eq:proof_prop_1_4b_Mt}
	\end{align}
	where we also used the fact that
	\begin{align*}
		\|\hat{w}^*_{t-j|t}-w_{t-j}\|_Q^2 \le 2\|\hat{w}^*_{t-j|t}\|_Q^2 + 2\|w_{t-j}\|_Q^2
	\end{align*}
	by Cauchy-Schwarz and Young's inequality.	
	Using the same inequalities, we have
	\begin{align}
	& \ \|\hat{x}_{t-M_t|t}^*-x_{t-M_t} \|_{P_2}^2 \nonumber \\
	=& \ \|\hat{x}_{t-M_t}-x_{t-M_t} + \hat{x}_{t-M_t|t}^* - \hat{x}_{t-M_t} \|_{P_2}^2 \nonumber \\
	\le& \  2\|\hat{x}_{t-M_t}-x_{t-M_t} \|_{P_2}^2 + 2 \|\hat{x}_{t-M_t|t}^* - \hat{x}_{t-M_t} \|_{P_2}^2. \label{eq:proof_prop_1_cauchy}
	\end{align}
	Inserting~\eqref{eq:proof_prop_1_cauchy} into~\eqref{eq:proof_prop_1_4b_Mt} results in
	\begin{align}
	&\ W_\delta(\hat{x}_t,x_t) \nonumber \\
	\stackrel{\eqref{eq:MHE_objective},\eqref{eq:MHE_IOSS_cost}}{\leq} & \ 2\eta^{M_t} \|\hat{x}_{t-M_t}-x_{t-M_t}\|_{P_2}^2 + 2\sum_{j=1}^{M_t}\eta^{j-1} \|w_{t-j}\|_Q^2\nonumber \\
	&+ V_{\mathrm{MHE}}(\hat{x}_{t-M_t|t}^*,\hat{w}_{\cdot|t}^*,\hat{y}_{\cdot|t}^*,t). \label{eq:W_delta_value_function_terminal_cost}
	\end{align}
	As final step, we use the fact that
	\begin{align*}
		\|\hat{x}_{t-M_t}-x_{t-M_t}\|_{P_2}^2 &\le \lambda_{\max}(P_2,P_1) \|\hat{x}_{t-M_t}-x_{t-M_t}\|_{P_1}^2 \\
		&\overset{\eqref{eq:dIOSS_quad_bounds}}{\le} \lambda_{\max}(P_2,P_1) W_\delta(\hat{x}_{t-M_t},x_{t-M_t}),
	\end{align*}
	Application to~\eqref{eq:W_delta_value_function_terminal_cost} yields the desired bound~\eqref{eq:W_delta_value_function}. 
\end{proof}

Applying the bounds on $W_\delta$ in~\eqref{eq:dIOSS_quad_bounds} and the definition of the estimation error in~\eqref{eq:MHE_estimate_error},
Inequality~\eqref{eq:W_delta_value_function} provides a bound on the estimation error $\hat{e}_t$ dependent on the value function, the past estimation error $\hat{e}_{t-M}$, and the past disturbances $w_{t-j}$ with $j\in\mathbb{I}_{\left[1,M_t\right]}$.
Hence, if we have uniform bounds on the past estimation error $\hat{e}_{t-M_t}$ and disturbances $w$, the value function provides a measure for the uncertainty in the state estimate, i.e., for a large value function, we have a large bound on the estimation error, and thus large uncertainty in the state estimate.
Consequently, the value function can serve as measure for the accuracy of the current state estimate.
Related estimation error bounds were derived for Luenberger-like observers and MHE in \cite[Prop. 1]{koehler2021Output} and \cite[Thm. 3]{koehler2021Output}, respectively.
We note that the result in~\cite{koehler2021Output} rely on a (local) continuity condition for $W_\delta$, while the presented result exploits quadratic bounds in~\eqref{eq:dIOSS_quad_bounds} to derive a linear bound in~\eqref{eq:W_delta_value_function}.
In the following, we use the bound derived in Proposition~\ref{prop:estimation_error_bound} to show that the $\delta$-IOSS Lyapunov function is an $M$-step Lyapunov function for MHE.

\begin{theorem}[$M$-step Lyapunov function for MHE] \label{thm:MHE_Lyapunov_function}
	Let Assumption~\ref{ass:IOSS_quadratically_bounded} hold.
	Then, for all $t\in\mathbb{I}_{\ge 0}$, the state estimate~$\hat{x}_t$ in~\eqref{eq:MHE_estimate} satisfies
	\begin{align}
		W_\delta(\hat{x}_t,x_t)	\leq&\ 4\eta^{M_t}\lambda_{\max}(P_2,P_1) W_\delta(\hat{x}_{t-M_t},x_{t-M_t}) \nonumber \\
		&\ +4\sum_{j=1}^{M_t}\eta^{j-1}\|w_{t-j}\|_Q^2. \label{eq:M-step_Lyap_decrease}
	\end{align}
\end{theorem}
\begin{proof}
	Since we assume the true underlying system to satisfy the constraints \eqref{eq:constraints}, the true disturbance, state, and output sequences are a feasible solution to the MHE problem~\eqref{eq:MHE_IOSS}, i.e., $V_{\mathrm{MHE}}(\hat{x}_{t-M|t}^*,\hat{w}_{\cdot|t}^*,\hat{y}_{\cdot|t}^*,t)\leq V_{\mathrm{MHE}}(x_{t-M_t},{w}_{\cdot|t},{y}_{\cdot|t},t)$ by optimality.
	Inserting this in Inequality\footnote{The resulting bound~\eqref{eq:M-step_Lyap_decrease} is identical if we directly use the bound~\eqref{eq:W_delta_value_function} from Proposition~\ref{prop:estimation_error_bound} here. However, the intermediate bound~\eqref{eq:M-step_bound_W_delta_prior_weighting} below will be used later in the proof of Corollary~\ref{cor:MHE_RGES}.}~\eqref{eq:W_delta_value_function_terminal_cost} and using~\eqref{eq:MHE_objective} results in
	\begin{align}
	&\ W_{\delta}(\hat{x}_t,x_t) \nonumber\\
	\leq&\ 4\eta^{M_t} \|\hat{x}_{t-M_t}-x_{t-M_t} \|_{P_2}^2 + 4\sum_{j=1}^{M_t}\eta^{j-1}\|w_{t-j}\|_Q^2. \label{eq:M-step_bound_W_delta_prior_weighting}
	\end{align}
	Using~\eqref{eq:dIOSS_quad_bounds} to upper bound $\|\hat{x}_{t-M_t}-x_{t-M_t} \|_{P_2}^2$ as in the proof of Proposition~\ref{prop:estimation_error_bound} leads to~\eqref{eq:M-step_Lyap_decrease}, which concludes the proof.
\end{proof}

In case the horizon length $M$ of the MHE problem~\eqref{eq:MHE_IOSS} is chosen such that
\begin{align}\label{eq:M_cond}
\rho^M \coloneqq 4\eta^{M}\lambda_{\max}(P_2,P_1) < 1,
\end{align}
with $\rho \in \left[0,1\right)$, 
then, for any $t\ge M$, the bound~\eqref{eq:M-step_Lyap_decrease} in Theorem~\ref{thm:MHE_Lyapunov_function} results in
\begin{equation*}
W_\delta(\hat{x}_t,x_t)	\leq \rho^M W_\delta(\hat{x}_{t-M},x_{t-M}) +4\sum_{j=1}^{M}\eta^{j-1}\|w_{t-j}\|_Q^2.
\end{equation*}
Consequently, $W_\delta(\hat{x}_t,x_t)$ is an $M$-step (incremental) Lyapunov function for the estimation error~\eqref{eq:MHE_estimate_error}.

By its very nature, the MHE computes an estimate $\hat{x}_t$ that depends on the past estimate $\hat{x}_{t-M_t}$ and a sequence of $M_t$ most recent measurements $y_{t-j}$ for $j\in\left[1,M_t\right]$.
As such, it is not surprising that the resulting Lyapunov function is defined over $M$ steps, as opposed to standard Lyapunov functions~\cite{angeli2002lyapunov}.
We note that~\cite{ahmadi2008non}, for autonomous systems $x_{t+1}=f(x_t)$, provides a method to compute a standard Lyapunov function based on an $M$-step decrease condition. 
However, in the considered MHE case, the dynamics also depend on the past $M_t$ measurements and thus a resulting Lyapunov function would also depend on those quantities.
In conclusion, the $M$-step Lyapunov function for MHE presented above follows naturally from the definition of the MHE, and is preferred for the following stability analysis, since a Lyapunov function with one-step decrease would not have a concise analytic expression.
\begin{remark}[Alternative $M$-step Lyapunov-like function]\label{rem:alternative_Lyapunov_function}
	We note that instead of the $\delta$-IOSS Lyapunov function $W_\delta(\hat{x}_t,x_t)$, an alternative Lyapunov-like function is naturally given by the weighted sum of a past $\delta$-IOSS Lyapunov function, the value function, and a sum of the past $M$ disturbances:
	\begin{align}
		V_{\delta}(t) \coloneqq &\ 2\eta^{M} \lambda_{\max}(P_2,P_1) W_\delta(\hat{x}_{t-M},x_{t-M}) \label{eq:alternative_lyapunov} \\
		&\ + V_{\mathrm{MHE}}(\hat{x}_{t-M|t}^*,\hat{w}_{\cdot|t}^*,\hat{y}_{\cdot|t}^*,t) +2\sum_{j=1}^{M}\eta^{j-1} \|w_{t-j}\|_Q^2, \nonumber
	\end{align}
	for $t\ge M$.
	In particular this function satisfies
	\begin{align*}
		&\ V_{\delta}(t+M) \\
		\stackrel{\eqref{eq:MHE_objective},\eqref{eq:dIOSS_quad_bounds}}{\le} &\ 4\eta^{M} \lambda_{\max}(P_2,P_1) W_\delta(\hat{x}_{t},x_{t})+ 4\sum_{j=1}^{M}\eta^{j-1} \|w_{t+M-j}\|_Q^2  \\
		\stackrel{\eqref{eq:W_delta_value_function}}{\le}&\ 4\eta^{M} \lambda_{\max}(P_2,P_1) V_\delta(t) +4\sum_{j=1}^{M}\eta^{j-1} \|w_{t+M-j}\|_Q^2,
	\end{align*}
	where we used the fact that the true state, disturbance, and output sequences are a feasible solution of~\eqref{eq:MHE_IOSS} at time step $t+M$, yielding an upper bound of the value function, compare also the proof of Theorem~\ref{thm:MHE_Lyapunov_function}.
	Consequently, $V_\delta(t)$ is an $M$-step Lyapunov-like function for MHE provided the horizon $M$ is chosen such that~\eqref{eq:M_cond} holds.
	In the case where no disturbances act on the system model~\eqref{eq:sys_w}, i.e., $w_t = 0$ for all $t \in \mathbb{I}_{\ge 0}$, the Lyapunov-like function~\eqref{eq:alternative_lyapunov} reduces to a weighted sum of the past $\delta$-IOSS Lyapunov function and the value function.
	This is particularly interesting due to the parallelism to output tracking MPC~\cite[Sec. 4.3]{allan2020lyapunov}, where the Lyapunov function is given by a weighted sum of the $\delta$-IOSS Lyapunov function and the value function of the MPC.
	In particular, in~\cite{allan2020lyapunov}, it is shown that a non-trivial Lyapunov-like function based on a sequence of augmented infinite horizon control problems can be used to study the stability of output tracking MPC and MHE, respectively. In contrast, the stability results for output tracking MPC in~\cite{Grimm2005model},\cite[Sec. 4.1]{koehler2021dynamic} directly provide a Lyapunov function given by a sum of the storage and value function based on a different analysis.
	In a similar spirit, the derived alternative Lyapunov-like function for MHE in~\eqref{eq:alternative_lyapunov} follows naturally from the \mbox{$\delta$-IOSS} Lyapunov function (Assumption~\ref{ass:IOSS_quadratically_bounded}) and the definition of the cost function~\eqref{eq:MHE_objective}, see Section~\ref{sec:discussion} below for a detailed comparison of our stability analysis with the Lyapunov-like functions from~\cite{allan2021FIE,allan2020lyapunov}.
\end{remark}

In the following, we show how RGES of MHE follows directly from~\eqref{eq:M-step_Lyap_decrease}.

\begin{corollary}[MHE is RGES]\label{cor:MHE_RGES}
	Let Assumption~\ref{ass:IOSS_quadratically_bounded} hold, and suppose the horizon $M$ satisfies Inequality~\eqref{eq:M_cond}.
	Then, for all $t \in \mathbb{I}_{\ge 0}$, the estimation error~\eqref{eq:MHE_estimate_error} satisfies 
	\begin{align}
		\|\hat{e}_t\|_{P_1} \le& \max \left\{\vphantom{\frac{4}{1-\sqrt[4]{\rho}}}4\sqrt{\rho}^{t}\|\hat{e}_0\|_{P_2},\right. \label{eq:error_bound_M_plus}\\
		&\left. \max_{j\in\mathbb{I}_{[0,t-1]}}\left\{\frac{4}{1-\sqrt[4]{\rho}}\sqrt[4]{\rho}^{j}\|w_{t-j-1}\|_Q\right\}\right\}, \nonumber
	\end{align}
	with $\rho$ as defined in~\eqref{eq:M_cond}, i.e., the MHE estimator~\eqref{eq:MHE_IOSS} is an RGES estimator according to Definition~\ref{def:RGAS}. 
\end{corollary}

\begin{proof}
	Consider some time $t = kM + l$, with unique $l\in \mathbb{I}_{\left[0,M-1\right]}$ and $k \in \mathbb{I}_{\ge 0}$.
	Using Inequality~\eqref{eq:M-step_bound_W_delta_prior_weighting} we get
	\begin{align}
	W_\delta(\hat{x}_l,x_l) \leq 4\eta^{l} \|\hat{x}_{0}-x_{0} \|_{P_2}^2 + 4\sum_{j=1}^{l}\eta^{j-1}\|w_{l-j}\|_Q^2.  \label{eq:MHE_FIE_bound}
	\end{align}
	Further, applying the bound~\eqref{eq:M-step_Lyap_decrease} $k$ times with $M_t=M$ and $\rho$ as defined in~\eqref{eq:M_cond}, we arrive at
	\begin{align}
	&\ W_\delta(\hat{x}_t,x_t) \nonumber \\
	\leq&\ \rho^{kM}W_\delta(\hat{x}_{l},x_{l}) +4\sum_{i=0}^{k-1}\rho^{iM}\sum_{j=1}^{M}\eta^{j-1}\|w_{t-iM-j}\|_Q^2 \nonumber \\
	\stackrel{\eqref{eq:MHE_FIE_bound}}{\leq}& \rho^{kM} \Bigg(4\eta^l\|\hat{e}_0\|_{P_2}^2 + 4\sum_{j=1}^{l}\eta^{j-1}\|w_{l-j}\|_Q^2 \Bigg) \nonumber \\
	&+4\sum_{i=0}^{k-1}\rho^{iM}\sum_{j=1}^{M}\eta^{j-1}\|w_{t-iM-j}\|_Q^2 \nonumber \\
	\leq&\ 4\rho^{t}\|\hat{e}_0\|_{P_2}^2 + 4 \sum_{q=0}^{t-1} \rho^{q}\|w_{t-q-1}\|_Q^2, \label{eq:MHE_RGES_sum_form}
	\end{align}
	where we used that $\eta \le \rho$ and $t=kM+l$.
	Applying the lower bound in~\eqref{eq:dIOSS_quad_bounds}, taking the square-root, and using the fact that $\sqrt{a+b} \le \sqrt{a} + \sqrt{b}$ for all $a,b \ge 0$,	we obtain
	\begin{align*}
		\|\hat{e}_t\|_{P_1} &\le \sqrt{W_{\delta}(\hat{x}_t,x_t)} \\
		&\le 2\sqrt{\rho}^{t}\|\hat{e}_0\|_{P_2} + 2 \sum_{q=0}^{t-1} \sqrt{\rho}^{q}\|w_{t-q-1}\|_Q.
	\end{align*}
	To arrive at the max formulation in~\eqref{eq:RGAS}, we use\footnote{A similar bound transferring sum-based into max-based expressions was recently used in the MHE literature, e.g., in \cite[Eq. (25)]{knuefer2020time}, \cite[Prop. 3.13]{allan2021FIE}.}
	\begin{align}
	\sum_{q=0}^{t-1} \sqrt{\rho}^{q}\|w_{t-q-1}&\|_Q^2 \le \sum_{q=0}^{t-1} \sqrt[4]{\rho}^{q}\max_{q\in\mathbb{I}_{[0,t-1]}}\left\{\sqrt[4]{\rho}^{q}\|w_{t-q-1}\|_Q^2\right\} \nonumber \\
	\le &\frac{1}{1-\sqrt[4]{\rho}} \max_{q\in\mathbb{I}_{[0,t-1]}}\left\{\sqrt[4]{\rho}^{q}\|w_{t-q-1}\|_Q^2\right\}. \label{eq:sum_to_max}
	\end{align}
	Inserting the bound on the sum above results in
	\begin{align*}
	\|\hat{e}_t\|_{P_1} \le&\ 2\sqrt{\rho}^{t}\|\hat{e}_0\|_{P_2} \\
	&\ + \max_{q\in\mathbb{I}_{[0,t-1]}}\left\{\frac{2}{1-\sqrt[4]{\rho}}\sqrt[4]{\rho}^{q}\|w_{t-q-1}\|_Q\right\},
	\end{align*}
	which results in~\eqref{eq:error_bound_M_plus} and concludes the proof.
\end{proof}

\begin{remark}
	\label{rem:M_cond_quadratic}
	In case Assumption~\ref{ass:IOSS_quadratically_bounded} holds with a quadratic $\delta$-IOSS Lyapunov function $W_\delta(x,\tilde{x})=\|x-\tilde{x}\|_P^2$, then $P_2 = P_1=P$ and thus condition \eqref{eq:M_cond} reduces to $4\eta^M < 1$.
\end{remark}

\subsection{Full Information Estimation}\label{sec:FIE}
In the following, we show how the analysis of MHE stability presented above can be extended to general $\delta$-IOSS Lyapunov functions according to Definition~\ref{def:IOSS_Lyap}, without requiring quadratic bounds and gains as done in Assumption~\ref{ass:IOSS_quadratically_bounded}, if we instead use a full information estimation (FIE) scheme.
In contrast to MHE, FIE considers all available past input and output measurements $(u,y)$ to obtain the current state estimate $\hat{x}_t$ at time $t$.
Thus, the size of the FIE problem is growing with time, and it is in general not computationally tractable.
However, FIE serves as a theoretical benchmark for optimization-based state estimation, especially because of the stability implication form FIE to MHE~\cite{hu2017robust,hu2021MHE,knuefer2021MHE}.
Here, the following FIE analysis particularly illustrates why exponential $\delta$-IOSS (Assumption~\eqref{ass:IOSS_quadratically_bounded}) was required in the case of MHE.

The objective of the considered FIE problem is to minimize the following cost function
\begin{align}
&V_{\mathrm{FIE}}(\hat{x}_{0|t},\hat{w}_{\cdot|t},\hat{y}_{\cdot|t},t) = \eta^{t}\alpha_2\left(2\|\hat{x}_{0|t}-\hat{x}_{0}\|\right) \label{eq:FIE_objective} \\
&\quad +\sum_{j=1}^{t}\eta^{j-1}\left(\sigma_w(2\|\hat{w}_{t-j|t}\|)+\sigma_y(\|\hat{y}_{t-j|t}-y_{t-j}\|) \right). \nonumber
\end{align}
The main difference to the MHE objective~\eqref{eq:MHE_objective} is the use of $\sigma_{w}$, $\sigma_{y}$, and $\alpha_2$ of the $\delta$-IOSS Lyapunov function~\eqref{eq:IOSS_Lyap}, instead of the quadratic functions as provided in Assumption~\ref{ass:IOSS_quadratically_bounded}.
The FIE state estimate is then obtained by solving the following problem with $M_t=t$ at each time step~$t$
\begin{subequations}\label{eq:FIE_IOSS}
	\begin{align}\label{eq:FIE_IOSS_cost}
	\min_{\hat{x}_{0|t},\hat{w}_{\cdot|t}}& V_{\mathrm{FIE}}(\hat{x}_{0|t},\hat{w}_{\cdot|t},\hat{y}_{\cdot|t},t) \\ \label{eq:FIE_IOSS_1}
	\text{s.t. }&\eqref{eq:MHE_IOSS_1},\eqref{eq:MHE_IOSS_2},\eqref{eq:MHE_IOSS_3},\eqref{eq:MHE_IOSS_4},
	\end{align}
\end{subequations}
with the optimal state estimate denoted as
\begin{align}\label{eq:FIE_estimate}
	\hat{x}_t = \hat{x}_{t|t}^*.
\end{align}
In the case of MHE, Proposition~\ref{prop:estimation_error_bound} showed the relation of the $\delta$-IOSS Lyapunov function and the value function~\eqref{eq:MHE_IOSS_cost}, and Theorem~\ref{thm:MHE_Lyapunov_function} allowed us to conclude that $W_{\delta}(\hat{x}_t,x_t)$ is an $M$-step Lyapunov function for MHE.
In the case of FIE without Assumption~\ref{ass:IOSS_quadratically_bounded}, a similar analysis allows us to derive the bound on $W_{\delta}(\hat{x}_t,x_t)$ as presented in the following proposition.
\begin{proposition}[Bound on $W_\delta(\hat{x}_t,x_t)$]
	\label{prop:bound_W_FIE}
	Let the system~\eqref{eq:sys_w} admit a $\delta$-IOSS Lyapunov function according to Definition~\ref{def:IOSS_Lyap}.
	Then, for all $t\in\mathbb{I}_{\ge 0}$, the state estimate~$\hat{x}_t$~\eqref{eq:FIE_estimate} satisfies
	\begin{align}
	W_\delta(\hat{x}_t,x_t)	\leq&\ 2\eta^{t} \alpha_2 \left(2 \alpha_1^{-1}(W_\delta(\hat{x}_{0},x_{0}))\right) \label{eq:FIE_W_delta_bound} \\
	&\ +2\sum_{j=1}^{t}\eta^{j-1}\sigma_w(2\|w_{t-j}\|). \nonumber
	\end{align}
\end{proposition}
\begin{proof}
	Using the monotone increase property of $\mathcal{K}$-functions and the weak triangular inequality~\cite{sontag1989smooth}, we obtain
	\begin{align*}
	\alpha_2(\|\hat{x}_{0|t}-x_{0}\|) &\le \alpha_2(\|\hat{x}_{0|t}-\hat{x}_{0}\|+\|x_{0}-\hat{x}_{0}\|) \\
	&\le \alpha_2(2\|\hat{x}_{0|t}-\hat{x}_{0}\|) + \alpha_2(2\|x_{0}-\hat{x}_{0}\|).
	\end{align*}
	Applying the bound~\eqref{eq:IOSS_Lyap_2} on $W_{\delta}(\hat{x}_t,x_t)$ for $t$ times and using this inequality, we arrive at
	\begin{align*}
	&W_{\delta}(\hat{x}_t,x_t) \le \eta^{t} \alpha_2(2\|\hat{x}_{0}-x_{0}\|) \\
	&\quad + V_{\mathrm{FIE}}(\hat{x}_{0|t}^*,\hat{w}_{\cdot|t}^*,\hat{y}_{\cdot|t}^*,t)  +\sum_{j=1}^{t}\eta^{j-1} \sigma_w(2\|w_{t-j}\|),
	\end{align*}
	analogous to Inequality~\eqref{eq:W_delta_value_function} in the proof of Proposition~\ref{prop:estimation_error_bound}.
	An upper bound for the value function can be computed by using the true sequence as a candidate, resulting in
	\begin{align}\label{eq:FIE_W_delta_bound_before_substitution}
	W_{\delta}(\hat{x}_t,x_t) \le 2\eta^{t} \alpha_2(2\|\hat{x}_{0}-x_{0}\|) +2\sum_{j=1}^{t}\eta^{j-1} \sigma_w(2\|w_{t-j}\|).
	\end{align}
	Finally, using the lower bound in~\eqref{eq:IOSS_Lyap_1} we have
	\begin{align*}
		\|\hat{x}_{0}-x_{0}\| \le \alpha_1^{-1}(W_\delta(\hat{x}_{0},x_{0}))
	\end{align*}
	which leads to~\eqref{eq:FIE_W_delta_bound} and concludes the proof.
\end{proof}

The bound presented in Proposition~\ref{prop:bound_W_FIE} allows us to show that the FIE scheme is an RGAS state estimator, as established in the following corollary.

\begin{corollary}[FIE is RGAS]
	Let the system~\eqref{eq:sys_w} admit a $\delta$-IOSS Lyapunov function according to Definition~\ref{def:IOSS_Lyap}. Then, the FIE estimator~\eqref{eq:FIE_IOSS} is an RGAS estimator according to Definition~\ref{def:RGAS}.
\end{corollary}
\begin{proof}
	We start from~\eqref{eq:FIE_W_delta_bound_before_substitution}. Using the lower bound in~\eqref{eq:IOSS_Lyap_1} and the weak triangular inequality leads to
	\begin{align*}
	\|\hat{x}_t-x_t\|
	\le&\ \alpha_1^{-1}\left(4 \eta^t \alpha_2(2\|\hat{x}_0 - x_0\|)\right) \\
	&\ + \alpha_1^{-1}\Bigg(4 \sum_{j=1}^{t}\eta^{j-1}\sigma_{w}(2\|w_{t-j}\|)\Bigg).
	\end{align*}
	To bound the second term above, we use the same procedure as in the proof of Corollary~\ref{cor:MHE_RGES} to transform the sum into a max term, resulting in
	\begin{align*}
	&\ \alpha_1^{-1}\Bigg(4 \sum_{j=1}^{t}\eta^{j-1}\sigma_{w}(2\|w_{t-j}\|)\Bigg) \\
	\qquad \stackrel{\eqref{eq:sum_to_max}}{\le}&\  \max_{i\in\mathbb{I}_{[0,t-1]}} \alpha_1^{-1}\left(\frac{4}{1-\sqrt{\eta}}\sqrt{\eta}^i\sigma_{w}(2\|w_{t-i-1}\|)  \right).
	\end{align*}
	Inserting the inequality above we obtain
	\begin{align*}
		&\ \|\hat{x}_t-x_t\| \\
		\le&\ \max\left\{\vphantom{\frac{4}{1-\sqrt{\eta}}}2\alpha_1^{-1}\left(4 \eta^t \alpha_2(2\|\hat{x}_0 - x_0\|)\right), \right. \\
		&\ \left. \max_{j\in\mathbb{I}_{[0,t-1]}}2\alpha_1^{-1}\left(\frac{4}{1-\sqrt{\eta}}\sqrt{\eta}^{j}\sigma_{w}(2\|w_{t-j-1}\|) \right) \right\},
	\end{align*}
	which shows that the FIE estimator satisfies the Definition~\ref{def:RGAS}, and thus concludes the proof.
\end{proof}

\begin{remark}
	\label{rem:exponential}
	By considering a quadratically bounded $\delta$-IOSS Lyapunov function (Assumption~\ref{ass:IOSS_quadratically_bounded}), we could derive a simple condition~\eqref{eq:M_cond} to provide an $M$-step Lyapunov function for MHE.
	In case the system admits a $\delta$-IOSS Lyapunov function according to Definition~\ref{def:IOSS_Lyap}, but Assumption~\eqref{ass:IOSS_quadratically_bounded} does not hold, i.e., the system~\eqref{eq:sys_w} does not admit a quadratically bounded $\delta$-IOSS Lyapunov function, and the MHE objective is replaced by~\eqref{eq:FIE_objective} defined over the past $M_t$ time steps, Inequality~\eqref{eq:M-step_Lyap_decrease} is replaced by
	\begin{align*}
	W_\delta(\hat{x}_t,x_t)	\leq&\ 2\eta^{M_t} \alpha_2 \left(2 \alpha_1^{-1}(W_\delta(\hat{x}_{t-M_t},x_{t-M_t}))\right) \\
	&\ +2\sum_{j=1}^{M_t}\eta^{j-1}\sigma_w(2\|w_{t-j}\|),
	\end{align*}
	compare the proof of Proposition~\ref{prop:bound_W_FIE}.
	For general $\alpha_1, \alpha_2 \in\mathcal{K}_{\infty}$, no minimal horizon length $M$ ensuring a decrease can be found.
	Instead, typically the minimal horizon depends explicitly on a bound on the estimation error at time $t-M_t$, resulting in non-global stability results.
	This issue was also discussed in more detail in~\cite[Sec. 5.5.3]{allan2020lyapunov}.
	In~\cite[Thm. 18]{knuefer2021MHE}, asymptotic stability of MHE is established under certain conditions, for which Assumption~\ref{ass:IOSS_quadratically_bounded} is sufficient, using a nonlinear contraction. 
\end{remark}
\subsection{Discussion}\label{sec:discussion}
Developing MHE schemes for nonlinear systems generally always requires balancing practical designs against valid theoretical guarantees.
For example, one may choose a very simple scheme involving a zero prior weighting and standard quadratic penalties, i.e., (\ref{eq:MHE_objective}) with $\eta=1$ and $P_2=0$, compare~\cite{michalska1995moving}.
However, since past data is completely neglected in the design, the system must in general be observable to ensure stability of MHE, and furthermore, large estimation horizons may be required to obtain a performance comparable to FIE, cf. \cite[Sec.~4.3.1]{rawlings2020model}.
Therefore, a non-zero prior weighting seems appropriate which, on the other hand, may require intricate conditions to ensure stable estimation, cf. \cite{rao2003constrained}.

Establishing MHE for general detectable nonlinear systems that follow a practical design, provide good theoretical guarantees, and require conditions that can be easily verified has also emerged as a major problem in the more recent literature, cf. \cite{muller2017nonlinear,hu2017robust,allan2019moving,knufer2018robust}.
In particular, robust stability of MHE could be established in \cite{muller2017nonlinear,allan2019moving} based on $\delta$-IOSS using a general cost function that permits standard quadratic penalties, however, the robustness bounds deteriorate with an increasing estimation horizon.
Such a behavior is counter-intuitive and undesired since one would naturally expect better estimation results if more information is taken into account.
This issue could be avoided using a modified cost: either by adding a max-term that penalizes the largest single disturbance as in \cite{muller2017nonlinear,hu2017robust}, or by using a specific cost structure satisfying the triangle inequality, cf. \cite{knufer2018robust}.
Thus, a trade-off between a standard quadratic cost function and good performance guarantees for MHE has arisen.
In contrast, we were able to resolve this conflict of objectives: we both consider standard quadratic penalties (except for time-discounting) and provide theoretical guarantees that improve as $M$ increases.
We point out that comparable stability results were achieved earlier by using more general time-discounting with $\mathcal{KL}$-functions~\cite{knuefer2021MHE} or without discounting using a Lyapunov-like function~\cite{allan2021FIE,allan2020lyapunov}.
The following discussion compares these two structurally different approaches to the framework presented in Section~\ref{sec:estimation}.

\subsubsection*{General time-discounting \cite{knuefer2021MHE}} 
The requirements of Theorem~\ref{thm:MHE_Lyapunov_function} for guaranteed robust stability of MHE are fundamentally the same as in \cite{knuefer2021MHE}, namely, that the detectability property of the system (given by $\delta$-IOSS) must be suitably related to the cost function used for MHE by employing additional time-discounting.
Consequently, the robustness bounds derived in Section~\ref{sec:estimation} are qualitatively comparable to those from \cite[Thm.~14]{knuefer2021MHE} for the special case of exponential stability.
However, we point out that under certain conditions, \cite[Thm.~14]{knuefer2021MHE} also implies an asymptotic stability result using a nonlinear contraction.
In contrast, we require exponential detectability (i.e., a quadratically bounded $\delta$-IOSS Lyapunov function \eqref{eq:dIOSS_quad_bounds}) to achieve linear contraction in the proof of Proposition~\ref{prop:estimation_error_bound}, cf. Remark~\ref{rem:exponential}.
Note that this is in line with most of the recent results on nonlinear MHE \cite{hu2021MHE,allan2019moving,hu2017robust,muller2017nonlinear,knufer2018robust}, where (local) exponential detectability is applied to achieve linear contraction over the estimation horizon, see also \cite[Prop.~1]{allan2019moving} and \cite[Lem.~1]{hu2021MHE}.

Whereas \cite{knuefer2021MHE,knufer2018robust,hu2021MHE,allan2019moving,hu2017robust,muller2017nonlinear} build their analysis on properties of certain $\mathcal{KL}$-functions, we employ a Lyapunov characterization in Section~\ref{sec:estimation}.
To the best of the authors' knowledge, a Lyapunov function for nonlinear MHE has been missing in the literature and therefore is an interesting contribution on its own.
Moreover, based on the proposed analysis and, in particular, a general reasoning in Lyapunov coordinates, further theoretical insights and practical improvements arise.
First, verifying the required detectability condition in order to guarantee stability of MHE is rather straightforward in our setup. 
This becomes immediately apparent in Section~\ref{sec:dIOSS}, where we give a simple condition for computing a $\delta$-IOSS Lyapunov function that satisfies Assumption~\ref{ass:IOSS_quadratically_bounded}. 
Even though this directly implies a traditional $\mathcal{KL}$-characterization of \mbox{$\delta$-IOSS} as mentioned in Remark~\ref{rem:iIOSS} below, we have a stronger relationship between the $\delta$-IOSS Lyapunov function and the MHE cost function.
Indeed, since we circumvent the additional step of calculating the respective $\mathcal{KL}$-functions, the tuning of the cost function becomes more easy and intuitive (cf. Remark~\ref{rem:cost}).
Moreover, note that arguing in Lyapunov coordinates generally allows for less restrictive conditions on the minimal horizon length for guaranteed RGES of MHE compared to, e.g., \cite[Thm.~14]{knuefer2021MHE}, which can be seen in Table~\ref{tab:horizons} and the discussion below.
Finally, a beneficial feature of Definition~\ref{def:IOSS_Lyap} is that even in the asymptotic case (i.e., where $\alpha_1,\alpha_2$ are arbitrary $\mathcal{K}_\infty$-functions in \eqref{eq:IOSS_Lyap_1}), we can still use an exponential decrease in \eqref{eq:IOSS_Lyap_2} (without loss of generality).
This is exploited in Section~\ref{sec:FIE} where we address asymptotic stability of FIE, which in the end allows for a much simpler and more intuitive tuning of the FIE cost function compared to, e.g., \cite[Ass.~1]{knuefer2021MHE} and \cite[Ass.~3]{hu2021MHE} using general $\mathcal{KL}$-function inequalities.

\subsubsection*{Lyapunov-like function framework \cite{allan2021FIE,allan2020lyapunov}}
We point out that a Lyapunov approach to stability of FIE already appeared in the literature; in particular, a Lyapunov-like function (termed a Q-function) was used in \cite{allan2019lyapunov} to establish nominal stability of FIE, and the results were extended in \cite{allan2021FIE} and \cite[Chap. 5]{allan2020lyapunov} to RGES and RGAS of FIE\footnote{
	Note that RGES of MHE was shown in \cite[Thm.~4.2]{allan2021FIE} assuming that the underlying FIE is RGES and provides a linear contraction over the horizon, i.e., without explicitly constructing a Q-function for MHE, compare \cite[Sec.~5.5.3]{allan2020lyapunov} and \cite[Sec.~4]{knuefer2021MHE}.
}, respectively.
However, the Q-function differs significantly from the Lyapunov function presented in this work, especially in its non-trivial structure utilizing two time arguments.
The key ingredient in \cite{allan2021FIE}, \cite[Chap. 5]{allan2020lyapunov} is a sequence of augmented infinite-horizon problems, each considering the first~$t$ disturbances and zero disturbances thereafter.
As a result, each of this infinite-horizon problems has finite disturbance sequences and therefore well-defined solutions.
Then, at any time $t\in\mathbb{I}_{\geq0}$, the respective infinite-horizon cost function is compared to the truncated finite-horizon cost function considering the partial time interval $\mathbb{I}_{[0,j-1]}$ for some $j \in \mathbb{I}_{[0,t]}$.
This procedure allows to establish one-step dissipation in $j$ for each $j \in \mathbb{I}_{[0,t]}$.
However, since the resulting function is only semidefinite, it needs to be combined with the $\delta$-IOSS Lyapunov function to finally create the desired Q-function.
Taking into account its different components, (``pessimistic''\footnote{
	Wording according to the discussion below \cite[Cor.~3.18]{allan2021FIE}; it is not distinguished between the influences of long past or recent disturbances.
}) Lyapunov-like bounds on the Q-function are established in \cite[Prop.~3.14]{allan2021FIE} for exponential, and in \cite[Sec. 5.3]{allan2020lyapunov} for asymptotic stability.
In this context, note that the lower bound of the Q-function is given by the lower bound of the $\delta$-IOSS Lyapunov function; the upper bound, however, requires an additional stabilizability assumption of certain structure, cf. \cite[Ass.~3.6]{allan2021FIE}, \cite[Ass.~5.14]{allan2020lyapunov}.

Our analysis, on the other hand, is much simpler in many respects, simply enabled by an additional discount factor in the cost function (\ref{eq:MHE_objective}), compare also Remark~\ref{rem:alternative_Lyapunov_function}.
Note that this corresponds to a \textit{fading memory} design, a concept that has been widely used in the literature for decades in many research areas (see, e.g., \cite{boyd1985fading,maass2000neural}), and which was previously exploited also in the context of state estimation, e.g., to deal with model errors in Kalman filter applications, cf. \cite{sorenson1971recursive}.
Within our framework, this results in a strong connection between our notion of detectability and the cost such that the $\delta$-IOSS Lyapunov function immediately provides a valid bound for FIE, cf. Proposition~\ref{prop:bound_W_FIE}, and consequently, directly serves as an $M$-step Lyapunov function for MHE, cf. Theorem~\ref{thm:MHE_Lyapunov_function}.
As a result, we are able to avoid many (potentially conservative) steps, excessive over-approximations, and additional conditions such as stabilizability in the analysis.
Interestingly, the conditions on the cost function in terms of compatibility with $\delta$-IOSS are fundamentally similar for all the results considered above (except for the time-discounting), see Remark~\ref{rem:cost}, \cite[Ass.~3.5]{allan2021FIE}, \cite[Ass. 5.13]{allan2020lyapunov}, and \cite[Ass.~1]{knuefer2021MHE}.

\subsubsection*{Conditions on the horizon length for RGES of MHE}
In the following, we compare the methods discussed above by means of their respective conditions on the horizon length for guaranteed RGES of MHE, which illustrates the general benefit of arguing in Lyapunov coordinates.
For a broader overview, we also consider \cite[Thm.~1]{allan2019moving}, i.e., MHE based on a $\mathcal{KL}$-function characterization of $\delta$-IOSS, but without a time-discounted objective function as in \cite{knuefer2021MHE}.
For a fair comparison, we choose the cost functions for  \cite{allan2021FIE,knuefer2021MHE,allan2019moving} such that the smallest possible horizon follows in each case; more specifically, we consider $b(s,t)=\beta(2s,t)$ according to \cite[Rem.~6]{knuefer2021MHE} and $V_\mathrm{p}(\chi,\bar{x})=\Vert\chi-\bar{x}\Vert^2$ in \cite[Ass.~3]{allan2019moving}.
Since the respective analysis is much more involved for \cite{allan2021FIE}, we consider only an ideal (strict) lower bound on the minimal horizon length which follows from a vanishing prior weighting ($\underline{c}_\mathrm{x}=\overline{c}_\mathrm{x}\rightarrow0$ in \cite[Ass.~3.4]{allan2021FIE}) and under a perfect stabilizability condition ($c_\mathrm{c}\rightarrow0$ in \cite[Ass.~3.6]{allan2021FIE}).
Provided that a $\delta$-IOSS Lyapunov function according to Assumption~\ref{ass:IOSS_quadratically_bounded} is given, Table~\ref{tab:horizons} shows for each case the resulting constants $C>0$ and $\mu\in(0,1)$ for the general contraction condition $C\mu^M<1$ used to establish RGES of MHE.
\begin{table}[t]
	\begin{threeparttable}
		\centering
		\caption{Contraction condition $C\mu^M<1$ for guaranteed RGES of MHE for the proposed framework compared to different methods from the literature.}
		\label{tab:horizons}
		\setlength\tabcolsep{8pt} 
		\begin{tabular*}{\columnwidth}{@{\extracolsep{\fill}}ccc}
			\toprule
			\ \ Result & $C$  & $\mu$ \\\midrule
			\ \ Proposed \eqref{eq:M_cond}
			& $4\lambda_{\max}(P_2,P_1)$
			& $\eta$ \\[0.5ex]
			\ \ \cite[Thm.~4.2]{allan2021FIE}
			& $> \sqrt{4c_2/c_1}$ 
			& $>\sqrt[4]{1-(1-\eta)c_1/(4c_2)}$ \\[0.5ex]
			\ \ \cite[Thm.~14]{knuefer2021MHE}
			& $8c_2/c_1$
			& $\eta$ \\[0.5ex]
			\ \ \cite[Thm.~1]{allan2019moving}
			& $3\sqrt{8c_2/c_1}$
			& $\sqrt{\eta}$ \\[0.5ex]\bottomrule
		\end{tabular*}	
		\begin{tablenotes}
			\footnotesize
			\item $c_1=\lambda_{\min}(P_1)$, $c_2=\lambda_{\max}(P_2)$
		\end{tablenotes}
	\end{threeparttable}
\end{table}
Solving the conditions for the minimal stabilizing horizon length by $M_{\min} = \lceil -\log C/\log\mu \rceil$ and applying standard properties of the logarithmic function, we arrive at the following conclusions.
First, under optimal choices of the cost functions in terms of the horizon length, the contraction conditions from \cite[Thm.~14]{knuefer2021MHE} and \cite[Thm.~1]{allan2019moving} are (except for the additional factor 3) very similar to each other, despite a structurally different MHE design and proof technique\footnote{
	For both \cite[Thm.~14]{knuefer2021MHE} and \cite[Thm.~1]{allan2019moving}, note that the factor $8$ in the second column of Table~\ref{tab:horizons} could be easily replaced by $4$ by a straightforward extension of the respective derivation.
	The additional factor $3$ appearing in the last row results from a max-based $\delta$-IOSS bound used in \cite[Thm.~1]{allan2019moving} (whereas \cite[Thm.~14]{knuefer2021MHE} allows for using a less-restrictive sum-based $\delta$-IOSS bound) and could also be avoided by a suitable modification.
	Nevertheless, the above conclusions also apply to these improved conditions in the practical case where $\lambda_{\max}(P_2)/\lambda_{\min}(P_1)>\lambda_{\max}(P_2,P_1)$, compare also the simulation example in Section~\ref{sec:num_1}.}.
Second, we can generally conclude that our approach provides the least conservative estimate on the minimal horizon length $M_{\min}$ for guaranteed RGES of MHE; to see this, recall that $\lambda_{\max}(P_2)/\lambda_{\min}(P_1)\geq\lambda_{\max}(P_2,P_1)$ for all $P_2\succeq P_1\succ0$ and observe that each constant $C,\mu$ in Table~\ref{tab:horizons} has its minimal value at $c_1=c_2$. 
This general fact is also observed in the numerical example in Section~\ref{sec:num_1}, where we compute the minimal stabilizing horizon length for each case (cf. Table~\ref{tab:horizons_num}).
As a side remark, we note that a direct consequence of the choices made in the proof of \cite[Prop.~3.15]{allan2021FIE} is that generally no better contraction rate than $\mu=\sqrt[4]{3/4}\approx0.93$ and hence no smaller horizon length than $M=10$ can be obtained using \cite[Thm.~4.2]{allan2021FIE}, even in case of $\eta=0$ in~\eqref{eq:dIOSS_quad_gains} and under the ideal setup considered above.

Overall, the proposed MHE framework employs a practical (fading memory) least-squares cost function and provides theoretical stability and robustness guarantees that improve as the horizon length $M$ increases.
In addition, the arguments directly extend to the non-exponential case if FIE is used.
In the next section, we establish a simple condition to compute a quadratically bounded $\delta$-IOSS Lyapunov function according to Assumption~\ref{ass:IOSS_quadratically_bounded} so that an MHE design with guaranteed robust exponential stability is directly obtained.

\section[Synthesis of a d-IOSS Lyapunov functions based on differential dynamics]{Synthesis of $\delta$-IOSS Lyapunov functions based on differential dynamics}
\label{sec:dIOSS} 

In the following, we provide a constructive and systematic approach to derive $\delta$-IOSS Lyapunov functions for system~\eqref{eq:sys_w} based on its differential dynamics.
This results in simple matrix inequality conditions involving the Jacobians of $f$ and $h$, which can be efficiently verified using standard tools such as sum-of-squares (SOS) optimization or linear parameter-varying (LPV) embeddings.
In combination with Section~\ref{sec:estimation}, we hence provide a sufficient condition that directly yields an explicit MHE scheme that is guaranteed to be RGES.

The differential analysis of nonlinear systems plays an important role as simple and intuitive tools from linear control theory become applicable; 
for example, by shifting the analysis of convergence between arbitrary system trajectories to the study of the linearizations along each trajectory, see, e.g., \cite{lohmiller1998contraction}.
Global properties can then be inferred by using tools from differential geometry -- typically based on a suitable Riemannian (or Finsler) metric under which the differential displacements decrease with the flow of the system. 
This universal concept offers wide applicability in the analysis (cf. \cite{manchester2014transverse,forni2013dissip,forni2014finslerLyap}) and control (cf. \cite{manchester2018robustCCM,manchester2017control,koelewijn2021control}) of nonlinear systems.
In addition, as pointed out in \cite{lohmiller1998contraction}, it is also naturally suitable for characterizing the convergence of observers (see, e.g., \cite{manchester2021observer,sanfelice2012metric,sanfelice2015convergence}), and consequently, also for characterizing $\delta$-IOSS.

To this end, throughout the following we assume that $f,h$ in~\eqref{eq:sys_w} are continuously differentiable.
The corresponding linearizations at a given point $(x,u,w) \in \mathbb{X}\times\mathbb{U}\times\mathbb{W}$ are then given by
\begin{align}\label{eq:dIOSS_ABCD}
	\begin{split}
	A=\frac{\partial f}{\partial x}(x,u,w), \qquad B=\frac{\partial f}{\partial w}(x,u,w),\\[1.5ex]
	C=\frac{\partial h}{\partial x}(x,u,w), \qquad D=\frac{\partial h}{\partial w}(x,u,w).
	\end{split}
\end{align}

Now consider an arbitrary change of coordinates $\bar{x}=\phi(x)$ with $\phi:\mathbb{R}^n\rightarrow\mathbb{R}^n$, which results in the equivalent system dynamics
\begin{subequations}\label{eq:dIOSS_sys_y}
	\begin{align}
		\bar{f}(\bar{x},u,w)&:=\phi(f(\phi^{-1}(\bar{x}),u,w)),\\
		\bar{h}(\bar{x},u,w)&:=h(\phi^{-1}(\bar{x}),u,w). \label{eq:dIOSS_sys_y2}
	\end{align}
\end{subequations}

\begin{assumption}[Coordinate transformation]\label{ass:dIOSS}
	There exists a diffeomorphism $\phi:\mathbb{R}^n\rightarrow \mathbb{R}^n$ such that $\bar{h}$~\eqref{eq:dIOSS_sys_y2} is affine in $(\bar{x},w)$, and $\partial\bar{h}/\partial \bar{x}_i=0$ for all $i=1,\ldots n-p$.
\end{assumption}

Provided that Assumption~\ref{ass:dIOSS} holds, the transformed dynamics~\eqref{eq:dIOSS_sys_y} are such that the output $\bar{h}$~\eqref{eq:dIOSS_sys_y2} depends affinely on a subset of the system state $\bar{x}$, which is similar to the class of systems considered in \cite{manchester2021observer}. 
Note that this is a fairly general setup covering several observability normal forms and therefore many physical models that admit a corresponding transformation, cf. \cite[Rem.~1]{manchester2021observer} and compare also \cite[Sec.~5.1]{nijmeijer1990nonlinear} for further details. 
Moreover, as we show in the following remark, the design of $\phi$ is particularly simple when a linear combination of the state is measured, which is the case in many practical applications, compare also the example systems in Section~\ref{sec:num}.

\begin{remark}[Coordinate transformation]\label{rem:trafo}
	In case the output function~\eqref{eq:sys_w_2} is given by $h(x)=Cx$ for some $C$, Assumption~\ref{ass:dIOSS} can be trivially satisfied using a linear change of coordinates $\bar{x}=\phi(x) = Tx$ with $T$ being a suitable non-singular transformation matrix.
	In particular, if $T$ is chosen such that $h(x)=Cx = CT^{-1}\bar{x} = (0,\bar{C})\cdot\bar{x}=\bar{h}(\bar{x})$ holds for some $\bar{C}\in\mathbb{R}^{p\times p}$, then it immediately follows that $\bar{h}$ is linear (and thus affine) in $\bar{x}$ with $\partial \bar{h}/\partial\bar{x}_i=0$ for all $i = 1,\ldots,n-p$.
\end{remark}

We partition the state $\bar{x}$ into two parts $\bar{x} = \left( \bar{x}_x^\top, \bar{x}_y^\top \right)^\top$  with $\bar{x}_x\in\mathbb{R}^{n-p}$ and $\bar{x}_y\in\mathbb{R}^p$.
Then, the following Theorem yields a quadratically bounded $\delta$-IOSS Lyapunov function according to Assumption~\ref{ass:IOSS_quadratically_bounded}.

\begin{theorem}[$\delta$-IOSS Lyapunov function]\label{thm:dIOSS_Lyap}
	Let Assumption~\ref{ass:dIOSS} hold and $P:\mathbb{R}^n\rightarrow\mathbb{R}^{n\times n}$ be such that
	\begin{equation}\label{eq:dIOSS_Px}
		P(x) = \frac{\partial\phi}{\partial x}(x)^\top
		\bar{P}(\phi(x))
		\frac{\partial\phi}{\partial x}(x)
	\end{equation}
	with
	\begin{equation}\label{eq:dIOSS_Pblock}
		\bar{P}(\bar{x}) = \begin{pmatrix} \bar{P}_x(\bar{x}_x) & 0\\ 0 & \bar{P}_y\end{pmatrix}
	\end{equation}
	for some $\bar{P}_x:\mathbb{R}^{n-p}\rightarrow\mathbb{R}^{(n-p)\times(n-p)}$ and $\bar{P}_y\in\mathbb{R}^{p\times p}$.
	Let $\mathbb{X}$ be weakly geodesically convex\footnote{
		Geodesic convexity is a natural generalization of convexity for sets to Riemannian manifolds, which reduces to convexity for the special case of constant metrics.
		For a formal definition, see, e.g., \cite[Def. 2.6]{sanfelice2012metric}.
	}, and $\mathbb{W}$ be convex.
	If there exist $\eta\in[0,1)$ and symmetric matrices $P_1,P_2\succ0$ and $Q,R\succeq0$ such that
	\begin{align}\label{eq:dIOSS_LMI_1}
		&\begin{pmatrix}
		A^\top P_+A-\eta P - C^\top RC
		& A^\top P_+B - {C}^\top RD \\
		B^\top P_+A - D^\top RC 
		& B^\top P_+B - Q - D^\top RD
		\end{pmatrix}
		\preceq 0
	\end{align}
	and
	\begin{equation}\label{eq:dIOSS_P1P2}
		P_1 \preceq P(x) \preceq P_2
	\end{equation}	
	hold for all $(x,u,w) \in \mathbb{X}\times\mathbb{U}\times\mathbb{W}$ with $P_+ = P(x^+)$, then there exists a quadratically bounded $\delta$-IOSS Lyapunov function $W_\delta$ that satisfies Assumption~\ref{ass:IOSS_quadratically_bounded}.
\end{theorem}
The proof of Theorem~\ref{thm:dIOSS_Lyap} employs several properties and arguments from Riemannian geometry and can be found in Appendix~\ref{app:dIOSS_proof_thm}.

We point out that for a fixed transformation $\phi$, conditions \eqref{eq:dIOSS_LMI_1}-\eqref{eq:dIOSS_P1P2} reduce to linear constraints that need to be verified over the full domain $\mathbb{X}\times\mathbb{U}\times\mathbb{W}$.
Computationally tractable sufficient conditions in terms of LMIs can then be obtained by using, e.g, LPV embeddings (see, e.g., \cite{koelewijn2021incremental}) or SOS relaxations, cf. \cite{parrilo2003semidefinite,wei2021contraction}.
In case $\phi$ is treated as a decision variable (which may be less restrictive due to this additional degree of freedom), the conditions of Theorem~\ref{thm:dIOSS_Lyap} can be reformulated as a convex optimization problem in a similar manner as in \cite{manchester2021observer}.

The following corollary of Theorem~\ref{thm:dIOSS_Lyap} provides even simpler conditions for the case where $h$ in~\eqref{eq:sys_w_2} is affine in $(x,w)$ and we restrict ourselves to a quadratic $\delta$-IOSS Lyapunov function.

\begin{corollary}[Quadratic $\delta$-IOSS Lyapunov function]\label{cor:dIOSS_Lyap}
	Let the output function $h$ in \eqref{eq:sys_w_2} be affine in $(x,w)$ and let $\mathbb{X}$ and $\mathbb{W}$ be convex.
	If there exist $\eta\in[0,1)$ and symmetric matrices $P\succ0$ and $Q,R\succeq0$ such that
	\begin{align}\label{eq:dIOSS_LMI_2}
		&\begin{pmatrix}
		A^\top PA-\eta P - C^\top RC
		& A^\top PB - {C}^\top RD \\
		B^\top PA - D^\top RC 
		& B^\top PB - Q - D^\top RD
		\end{pmatrix}
		\preceq 0
	\end{align}
	holds for all $(x,u,w) \in \mathbb{X}\times\mathbb{U}\times\mathbb{W}$, then $W(x,\tilde{x}) = \|x - \tilde{x}\|_P^2$ is a $\delta$-IOSS Lyapunov function and satisfies Assumption~\ref{ass:IOSS_quadratically_bounded} with $P_1 = P_2 = P$.
\end{corollary}

The proof of Corollary~\ref{thm:MHE_Lyapunov_function} is provided in Appendix~\ref{app:dIOSS_proof_cor}.
Some remarks are in order.

\begin{remark}[Relation to dissipativity]
	The proof of Theorem~\ref{thm:dIOSS_Lyap} introduces a differential version of IOSS (compare Claim~\ref{claim:dIOSS_1} in Appendix~\ref{app:dIOSS_proof_thm} for further details).
	This characterization is equivalent to the notion of differential (Q,S,R)-dissipativity \cite[Def.~3]{verhoek2020convex} with $S=0$, compare also \cite{koelewijn2021incremental,forni2013dissip}.
	However, as pointed out in \cite[Rem. 7]{verhoek2020convex}, the corresponding works crucially rely on $R\preceq0$ in order to derive incremental results by simply exhausting the Cauchy-Schwartz inequality, cf. \cite[Lem. 16]{koelewijn2021incremental}, compare also \cite[Thm. 1]{manchester2018robustCCM} and \cite[Thm. 2.4]{wei2021contraction}.
	Note that in our case, this would restrict the results to open-loop stable systems (since \eqref{eq:dIOSS_quad_gains} would need to hold with $R=0$, which directly implies $\delta$-ISS of system \eqref{eq:sys_w}). Moreover, this would result in the cost function~\eqref{eq:MHE_objective} not being positive definite, which generally can lead to an ill-defined optimization problem~\eqref{eq:MHE_IOSS}.
	In contrast, we circumvent this technical condition by suitably relating the state and output manifolds as it was similarly done in \cite{sanfelice2021metric,manchester2021observer} for observer design.
	More specifically, from Assumption~\ref{ass:dIOSS}, i.e., by imposing the existence of coordinates $\bar{x}$ in which the output function $\bar{h}$ is affine (which directly implies that $\bar{h}$ is totally geodesic by assumption, cf. \cite{vilms1970totally}), and due to our choice of the metric $\bar{P}(\bar{x})$ according to Theorem~\ref{thm:dIOSS_Lyap} (or Corollary~\ref{cor:dIOSS_Lyap}), we immediately obtain an equality relation between the integral of the differential supply rates and the incremental supply rates, compare \eqref{eq:dIOSS_int_w} and \eqref{eq:dIOSS_int_y} in Appendix~\ref{app:dIOSS_proof_thm} for details.
	Consequently, as a side result, we note that Theorem~\ref{thm:dIOSS_Lyap} (and Corollary~\ref{cor:dIOSS_Lyap}) with $\eta=1$ can be used to verify incremental dissipativity of system \eqref{eq:sys_w} subject to a positive definite supply rate, relaxing \cite[Rem. 7]{verhoek2020convex}.
\end{remark}

\begin{remark}[Extensions]\label{rem:extensions}
	To further generalize the pa\-ram\-e\-triza\-tion of $\bar{P}(\bar{x})$ with respect to $\bar{x}$, we note that the following minor extension of Theorem~\ref{thm:dIOSS_Lyap} is possible if, e.g., $\bar{h}(\bar{x}) = \bar{x}_y$ (neglecting $u$ and $w$ for ease of presentation).
	We could choose
	\begin{equation}\label{eq:dIOSS_P_extension}
		\bar{P}(\bar{x}) = \begin{pmatrix} \bar{P}_x(\bar{x}_x) & 0\\ 0 & \bar{P}_y(\bar{x}_y)\end{pmatrix}
	\end{equation}
	with $\bar{P}_{y,1}\preceq\bar{P}_y(\bar{x}_y)\preceq\bar{P}_{y,2}$ uniformly for all possible $\bar{x}_y$ and some constant matrices $\bar{P}_{y,1},\bar{P}_{y,2}\succ0$; i.e., $\bar{P}$ in~\eqref{eq:dIOSS_Pblock} with an additional dependency of $\bar{P}_y$ on $\bar{x}_y$.
	Then, by additionally imposing $R \preceq \bar{P}_y(\bar x_y)$ in~\eqref{eq:dIOSS_LMI_1}, one can derive a similar $\delta$-IOSS Lyapunov function as provided by Theorem~\ref{thm:dIOSS_Lyap} that satisfies Assumption~\ref{ass:IOSS_quadratically_bounded}; the technical details can be found in Remark~\ref{rem:extensions_details} in Appendix~\ref{app:dIOSS_proof_thm}.
	Finally, we note that one may relax Assumption~\ref{ass:dIOSS}, i.e., affinity of $\bar{h}$, by imposing that $\bar{h}$ is a Riemannian submersion,  cf.\cite{sanfelice2021metric}.		
\end{remark}

\begin{remark}[Closed-form expression] \label{rem:iIOSS}
	Note that Theorem~\ref{thm:dIOSS_Lyap} yields only an implicit $\delta$-IOSS Lyapunov function $W_\delta$, which is due to the fact that we have no analytical closed-form expression for the Riemannian energy of the minimizing geodesic, compare Appendix~\ref{app:dIOSS_proof_thm} for further details.
	However, note also that this is not needed for the particular MHE (or FIE) scheme, since we only require knowledge of the matrices $P_2,Q,R$ to design the cost functions~\eqref{eq:MHE_objective} and \eqref{eq:FIE_objective}, and additionally $P_1$ to compute the minimal horizon length for guaranteed RGES of MHE, cf. Section~\ref{sec:estimation}.
	Similar considerations apply if Theorem~\ref{thm:dIOSS_Lyap} is used to compute the $\mathcal{KL}$-functions of the standard $\delta$-IOSS bound. Again, one only needs to know the matrices $P_1,P_2,Q,R$ and use \eqref{eq:dIOSS_quad_bounds} after repeated application of the dissipation inequality~\eqref{eq:dIOSS_quad_gains} to obtain the desired result.
	If nevertheless an analytical expression for the $\delta$-IOSS Lyapunov function $W_\delta$ is desired, Corollary~\ref{cor:dIOSS_Lyap} can be used to obtain a quadratic function.
\end{remark}

\begin{remark}[Alternative derivation]
	\label{rem:alternative_W}
	An alternative way to compute a quadratic \mbox{$\delta$-IOSS} Lyapunov function is to first design an RGES observer based on, e.g., \cite{sanfelice2015convergence,manchester2021observer,astolfi2022constrained}.
	Then, under certain conditions, one can show that the corresponding Lyapunov function also serves as a $\delta$-IOSS Lyapunov function, cf. \cite[Prop.~4]{koehler2021Output}, and compare also \cite[Sec.~VII]{schiller2021suboptimal}.
	However, these sufficient conditions are crucially limited to quadratic Lyapunov functions and additive disturbances in the dynamics~\eqref{eq:sys_w_1}, and hence are only applicable to a smaller class of detectable systems (in comparison to Theorem~\ref{thm:dIOSS_Lyap}).
\end{remark}
\section{Numerical examples}
\label{sec:num}

In order to illustrate our results, we apply the proposed methods to two examples from the literature: a chemical reactor process (cf. Section~\ref{sec:num_1}) and a 12-state quadrotor model with flexible rotor blades (cf. Section~\ref{sec:num_2}).
The simulations were performed in Matlab using CasADi \cite{andersson2019casadi} and the NLP solver IPOPT \cite{waechter2005ipopt}; LMIs were verified using YALMIP \cite{loefberg2009sos} and the semidefinite programming solver MOSEK \cite{mosek2019}.

Overall, these examples demonstrate the practicability of the offline $\delta$-IOSS verification (Theorem~\ref{thm:dIOSS_Lyap}), the (significantly) shorter horizon bounds obtained through Theorem~\ref{thm:MHE_Lyapunov_function} compared to the literature (cf. Table~\ref{tab:horizons_num} below), and the applicability of the proposed MHE framework---in particular, its ability to provide valid theoretical guarantees under practical conditions.
		
\subsection{2-state chemical reaction}\label{sec:num_1}
\begin{figure}
	\centering
	\includegraphics{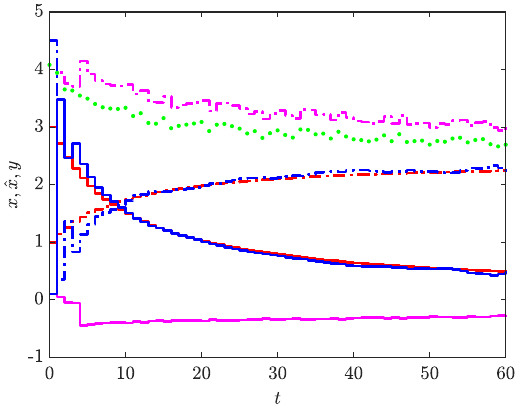}
	\caption{Comparison of MHE results (blue), EKF estimates (magenta), real system states (red) and measurements (green circles) for the chemical reaction, where $x_1,\hat{x}_1$ are solid and $x_2,\hat{x}_2$ are dash-dotted. We have used $M{\,=\,}30$, which satisfies condition~\eqref{eq:M_cond} and hence guarantees RGES of MHE by Theorem~\ref{thm:MHE_Lyapunov_function}.}
	\label{fig:1}
\end{figure}

We consider the following system 
\begin{align*}
	x_1^+ &= x_1 + t_{\Delta}(-2k_1x_1^2+2k_2x_2) + w_1,\\
	x_2^+ &= x_2 + t_{\Delta}(k_1x_1^2-k_2x_2) + w_2,\\
	y &= x_1 + x_2+ w_3,
\end{align*}
with $k_1=0.16$, $k_2=0.0064$, and sampling time $t_{\Delta}=0.1$. 
This corresponds to the chemical reaction $2A \rightleftharpoons B$ taking place in a constant-volume batch reactor from~\cite[Sec.~5]{tenny2002efficient} using an Euler discretization and with additional disturbances~$w\in\mathbb{R}^3$.
In the following, we treat $w$ as a uniformly distributed random variable satisfying $|w_i| \leq 10^{-3}, i =  1, 2$ for the process disturbances and $|w_3| \leq 0.1$ for the measurement noise.
As in~\cite[Sec.~5]{tenny2002efficient}, we consider $x_0=[3,1]^\top$ and the poor initial estimate $\hat{x}_0=[0.1,4.5]^\top$.
This setup poses a challenge for state estimators; in fact, simple estimators such as the standard extended Kalman filter (EKF) can fail to provide meaningful results, compare the simulation results in Figure~\ref{fig:1}.
This example is also frequently used in the related MHE literature (e.g., \cite[Example~4.38]{rawlings2020model}), however, $\delta$-IOSS has never been certified.

To this end, we assume that the prior knowledge $\mathbb{X} = [0.1,4.5]\times[0.1,4.5]$ is available, which follows from the physical nature of the system under the above conditions (in particular, the initial conditions and boundedness of $\mathbb{W}$), compare also the simulation results in Figure~\ref{fig:1} below.
For the considered system, we can even apply Corollary~\ref{cor:dIOSS_Lyap} in combination with SOS optimization to compute a quadratic Lyapunov Function $W_\delta=\Vert x-\tilde{x}\Vert_P^2$ that satisfies Assumption~\ref{ass:IOSS_quadratically_bounded} with
\begin{equation*}
	P = \begin{bmatrix}  4.539 & 4.171 \\ 4.171 & 3.834 \end{bmatrix},\
	Q = \begin{bmatrix}  10^3 & 0 & 0 \\ 0 & 10^4 & 0 \\ 0 & 0 & 10^3 \end{bmatrix},\
	R = 10^3,
\end{equation*}
and the decay rate $\eta=0.91$.
We point out that, to the best of the authors' knowledge, this is the first time that $\delta$-IOSS has been explicitly verified for this example.
It is also worth noting that the lack of such a method in the literature was generally considered a major problem in~\cite{allan2021FIE}, since $\delta$-IOSS became a standard detectability assumption in the recent nonlinear MHE literature, compare Section~\ref{sec:discussion}.
Theorem~\ref{thm:dIOSS_Lyap} provides a useful tool to actually verify this crucial property in practice.

Based on the $\delta$-IOSS Lyapunov function above, we can now compute the minimum horizon length $M_{\min}$ sufficient for robust stability of MHE according to condition~\eqref{eq:M_cond} and Remark~\ref{rem:M_cond_quadratic}, and compare it to corresponding bounds from the recent nonlinear MHE literature, i.e., the Lyapunov-like function framework \cite{allan2021FIE}, MHE with general time-discounting \cite{knuefer2021MHE}, and without time-discounting \cite{allan2019moving}, by resolving the respective conditions in Table~\ref{tab:horizons}.
As can be seen from Table~\ref{tab:horizons_num}, the proposed Lyapunov approach yields a minimum horizon length that is (at least) one order of magnitude better (i.e., smaller) than those obtained from the literature.
\begin{table}
	\centering
	\caption{Minimum required horizon length for guaranteed RGES of MHE compared to the different methods from the literature considered in Table~\ref{tab:horizons}.}
	\setlength\tabcolsep{5pt} 
	\begin{tabular*}{\columnwidth}{@{\extracolsep{\fill}}ccccc}	
		\toprule
		\ \ Result & Proposed \eqref{eq:M_cond} & \cite[Thm. 4.2]{allan2021FIE} & \cite[Thm. 14]{knuefer2021MHE} & \cite[Thm. 1]{allan2019moving} \\\midrule
		\ \ $M_{\min}$ & $15$ & $>8\cdot10^6$ & $119$ & $142$ \\\bottomrule
	\end{tabular*}	
	\label{tab:horizons_num}
\end{table}

For the following simulation, we choose $M=30>M_{\min}$ to provide a small estimation error bound.
The simulation results are depicted in Figure~\ref{fig:1}, which shows robustly stable estimation as guaranteed by Theorem~\ref{thm:MHE_Lyapunov_function} due to satisfaction of condition~\eqref{eq:M_cond}.
In order to compare the results, we also simulated the EKF. As can be seen in Figure~\ref{fig:1}, however, the corresponding estimates exhibit a serious error compared to MHE, which is partly due to the fact that the physical constraints were not met.
In summary, the overall simulation results are similar to \cite[Sec.~5]{tenny2002efficient}, \cite[Example~4.38]{rawlings2020model}, but with valid robustness guarantees for MHE.

\subsection{12-state quadrotor model}\label{sec:num_2}

We adapt the example from \cite{kai2017quadrotor} and consider a quadrotor model involving four rotors with flexible blades.
Let $\mathcal{I}$ denote the stationary inertial system with its vertical component pointing into the Earth, where position and velocity of the quadrotor are represented by $z=[z_1,z_2,z_3]^\top$ and $v=[v_1,v_2,v_3]^\top$, respectively.
By $\mathcal{B}$ we denote the body-fixed frame attached to the quadrotor, with the third component pointing in the opposite direction of thrust generation.
The attitude of $\mathcal{B}$ with respect to $\mathcal{I}$ is captured by a rotation matrix $R$ (where we use $zyx$-convention), which involves the roll, pitch, and yaw angle of the quadrotor represented by $\xi=[\phi,\theta,\psi]^\top$.
The angular velocity of the quadrotor in $\mathcal{B}$ with respect to $\mathcal{I}$ is given by $\Omega = [\Omega_1,\Omega_2,\Omega_3]^\top$.
Assuming a wind-free environment, the overall dynamics can be described as
\begingroup
\renewcommand*{\arraystretch}{1.2}
\begin{align*}
\begin{matrix*}[l]
\dot{z}=v, &m\dot{v}=mge_3-TR(\xi)e_3-R(\xi)B\Omega,\\
\dot{\xi}=\Gamma(\xi)\Omega, &J\dot{\Omega} = -\Omega^{\times}J\Omega+\tau-D\Omega,
\end{matrix*}
\end{align*}
\endgroup
where $e_3 = [0\ 0\ 1]^\top$ and $(\cdot)^{\times}$ refers to the skew symmetric matrix associated with the cross product such that $u^{\times}v = u\times v$ for any $u,v\in\mathbb{R}^3$.
The thrust $T\in\mathbb{R}$ and the torque $\tau\in\mathbb{R}^3$ are generated by the four rotors by means of their angular velocities $\omega_i$ via
$\begin{bmatrix} T \\ \tau \end{bmatrix} = 
	\text{\small$
	\begin{bmatrix}
		c_T & c_T & c_T & c_T\\
		0 & -lc_T & 0 & lc_T\\
		lc_T & 0 & -lc_T & 0\\
		-c_Q & c_Q & -c_Q & c_Q
	\end{bmatrix}$}
	\text{\footnotesize$\begin{bmatrix} \omega_1^2 \\ \omega_2^2 \\ \omega_3^2 \\ \omega_4^2 \end{bmatrix}$},$
and the matrix $\Gamma$ is defined as
$\Gamma(\xi) = \text{$
	\begin{bmatrix}
		1 & \sin\phi\tan\theta & \cos\phi\tan\theta\\
		0 & \cos\phi & -\sin\phi\\
		0 & \sin\phi\sec\theta & \cos\phi\sec\theta
	\end{bmatrix}$},$
compare~\cite{kai2017quadrotor} and\cite{nascimento2019position} for further details on the model and its derivation.
The parameters are chosen as $m=1.9$, $J = \mathrm{diag}(5.9, 5.9, 10.7){\,\cdot\,}10^{-3}$, $g=9.8$, $l=0.25$, $c_T=10^{-5}$, $c_Q = 10^{-6}$, $B = 1.14\cdot e_3^{\times}$, and $D = 0.0297 \cdot e_3e_3^\top$.
In summary, the overall model has the states $x=[z^\top \, v^\top \, \xi^\top \, \Omega^\top]^\top\in\mathbb{R}^{12}$ and the inputs $u = [\omega_1 \ \omega_2 \ \omega_3 \ \omega_4]^\top\in\mathbb{R}^4$.
We additionally assume that the dynamics of $\dot{x}_i$ is corrupted by an additive disturbance $d_i$, $i\in\mathbb{I}_{[1,12]}$, and that only noisy position and orientation measurements $y=[z^\top\, \xi^\top]^\top+v$ with noise $v\in\mathbb{R}^{6}$ are available.
In the following, we consider $d,v$ uniformly distributed such that $|d_i|\leq10^{-3}$, $i=\mathbb{I}_{[1,12]}$, and $|v_i|\leq0.1$, $i=\mathbb{I}_{[1,6]}$ and define $w = [d^\top\,v^{\top}]^{\top}\in\mathbb{R}^{18}$.
The discrete-time model~\eqref{eq:sys_w} is then obtained via Euler-discretization using the sampling time $t_\Delta = 0.05$.

We assume that some input/output sequences $(u,y)$ have been measured while performing a certain control scenario of the quadrotor that guarantees $x\in\mathbb{X} = \{x : |\xi_i|\leq\pi/6, |\Omega_i|\leq1, i\in\mathbb{I}_{[1,3]}\}$ and $u\in\mathbb{U} = \{u : |u_i|\leq1500, i\in\mathbb{I}_{[1,4]} \}$; the objective is to reconstruct the corresponding state trajectory using the proposed MHE framework. To this end, we verify condition~\eqref{eq:dIOSS_LMI_2} on $\mathbb{X}\times\mathbb{U}$ by suitably gridding the state space and thus compute a quadratic $\delta$-IOSS Lyapunov function with the decay rate $\eta=0.87$.
Choosing the horizon length $M=30$ satisfies condition~\eqref{eq:M_cond}, so that the proposed MHE design \eqref{eq:MHE_objective} and \eqref{eq:MHE_IOSS} is guaranteed to be RGES according to Theorem~\ref{thm:MHE_Lyapunov_function}.

\begin{figure}
	\centering
	\includegraphics{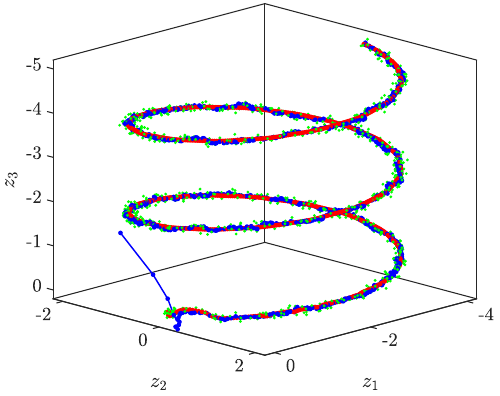}
	\caption{Comparison of the estimated (blue), true (red), and measured (green) position of the quadrotor.}
	\label{fig:quad_traj}
\end{figure}

\begin{figure}
	\centering
	\includegraphics{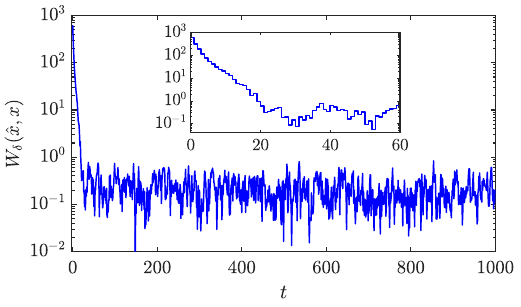}
	\caption{Estimation error of the quadrotor in Lyapunov coordinates.}
	\label{fig:quad_error}
\end{figure}

Figure~\ref{fig:quad_traj} shows the real, measured, and estimated position of the quadrotor (in the frame $\mathcal{I}$) to illustrate the maneuver flown.
The overall estimation error in Lyapunov coordinates is depicted in Figure~\ref{fig:quad_error} and illustrates exponential convergence to a neighborhood around the origin, as guaranteed by Theorem~\ref{thm:MHE_Lyapunov_function}.

\section{Conclusion}
\label{sec:sum} 
 
In this paper, we have presented a novel robust stability analysis for moving horizon estimation using Lyapunov functions.
The analysis generally applies to nonlinear exponentially detectable ($\delta$-IOSS) systems admitting a corresponding \mbox{$\delta$-IOSS} Lyapunov function.
Considering an MHE formulation with time-discounted quadratic objective, we have shown that an $M$-step Lyapunov function naturally arises, which directly implies robust exponential stability of MHE provided that the horizon length $M$ satisfies a posed lower bound.

The main feature of the proposed analysis is that, in contrast to most of the MHE literature, we argue entirely in Lyapunov coordinates; this is beneficial in several respects:
First, tuning the MHE objective by suitably relating it to $\delta$-IOSS in order to achieve good theoretical guarantees (which typically yields general $\mathcal{KL}$-function inequalities in the literature) becomes easy and intuitive, even when we show robust asymptotic stability for FIE;
second, the proposed Lyapunov analysis generally allows for less conservative (i.e., shorter) horizon bounds compared to recently proposed MHE designs.

Nonlinear detectability ($\delta$-IOSS) is a common detectability assumption in most of the existing recent results on nonlinear MHE; however, there was no systematic method of verifying this crucial condition so far, which was also considered a major problem in \cite{allan2021FIE} to establish guarantees for MHE beyond conceptual nature.
We were able to solve this issue by providing a systematic tool to verify $\delta$-IOSS for a large class of nonlinear detectable systems based on their differential dynamics.
The sufficient conditions were stated in terms of simple matrix inequalities that can be efficiently verified using, e.g., SOS optimization or LPV embeddings.
In combination, these conditions directly yield an MHE design with guaranteed robust exponential stability.

The applicability of the overall framework was illustrated with two examples from the literature: a standard MHE benchmark example where we verified $\delta$-IOSS for the first time, and a nonlinear 12-state quadrotor model.
In the end, we were able to achieve guaranteed robustly stable estimation under practical conditions enabled by a significantly shorter bound on the horizon length compared to the literature.

An interesting question for future research is under which conditions the MHE stability analysis proposed in Section~\ref{sec:MHE_stability} is also applicable in the case of a (relaxed) asymptotic $\delta$-IOSS condition or an objective without time-discounting.
Scalability of the approach to verify the underlying $\delta$-IOSS condition as presented in Section~\ref{sec:dIOSS} to higher-dimensional systems is mainly limted by the tools applied to verify the underlying matrix inequalities.

%


\begin{IEEEbiography}
	[{\includegraphics[width=1in,height=1.25in,clip,keepaspectratio]{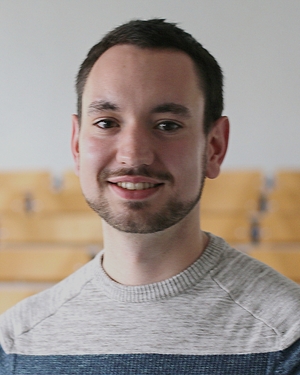}}]{Julian D. Schiller}
	received his Master degree in Mechatronics from the Leibniz University Hannover, Germany, in 2019. 
	Since then, he has been a research assistant at the Institute of Automatic Control, Leibniz University Hannover, where he is currently working on his Ph.D. under the supervision of Prof. Matthias A. Müller. 
	His research interests are in the area of optimization-based state estimation and the control of nonlinear systems.
\end{IEEEbiography}

\begin{IEEEbiography}
	[{\includegraphics[width=1in,height=1.25in,clip,keepaspectratio]{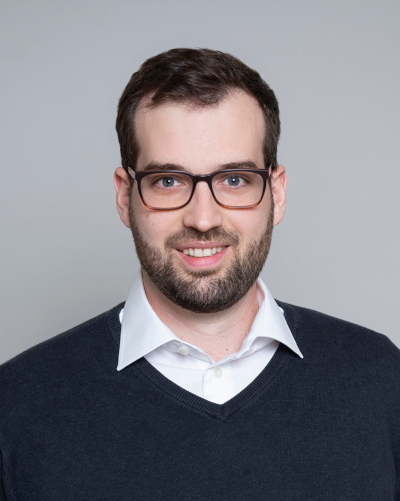}}]{Simon Muntwiler}
	received his Master degree in Robotics, Systems and Control from ETH Z\"urich, Switzerland, in 2019.
	He is currently a doctoral student at the Institute for Dynamic Systems and Control (IDSC) at ETH Z\"urich under the supervision of Prof. Melanie N. Zeilinger.
	His research interests are in the area of optimization- and learning-based state estimation and control algorithms, with application to safety critical systems.
\end{IEEEbiography}

\begin{IEEEbiography}
	[{\includegraphics[width=1in,height=1.25in,clip,keepaspectratio]{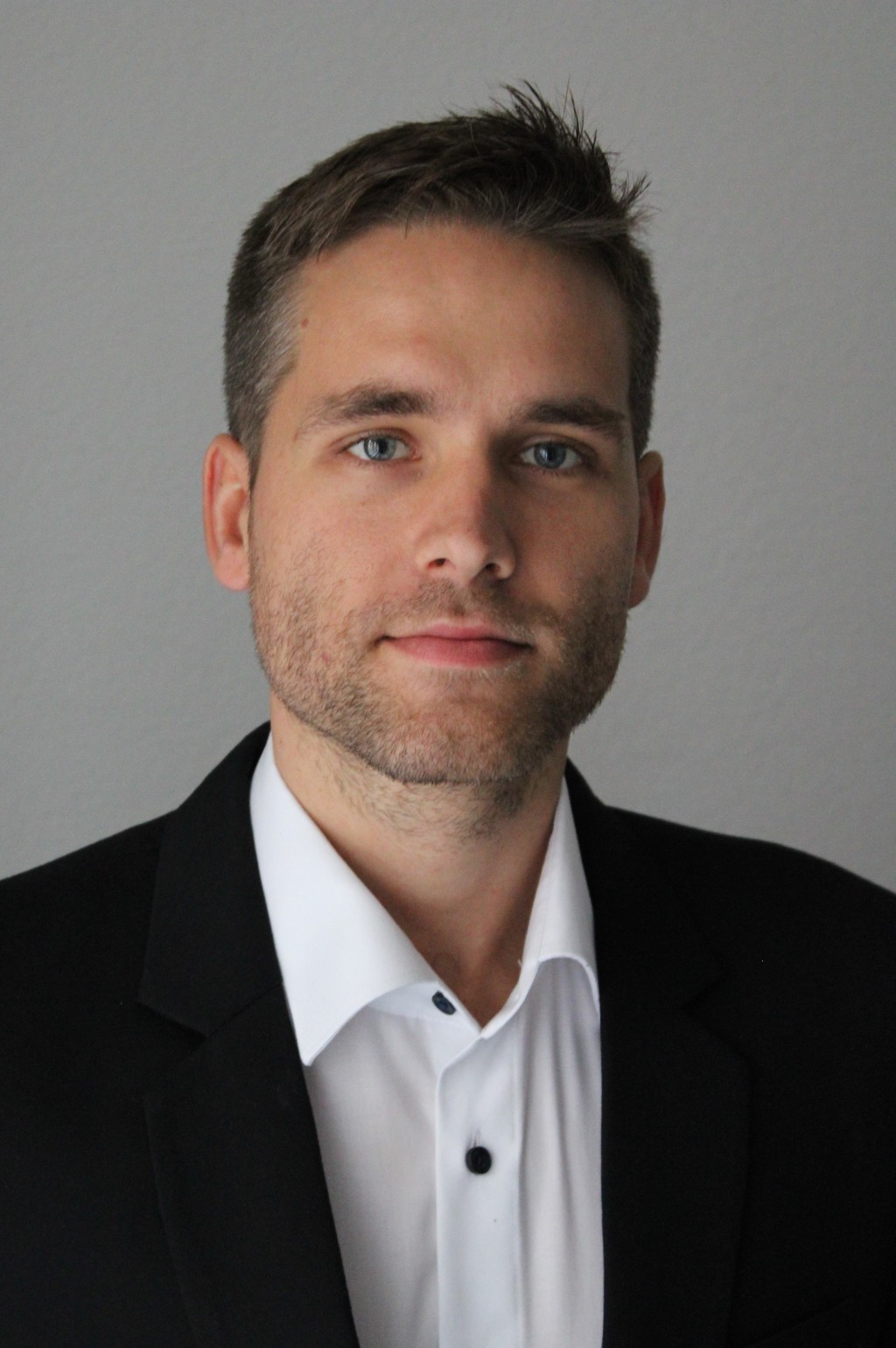}}]{Johannes K\"ohler}
	received his Master degree in Engineering Cybernetics from the University of Stuttgart, Germany, in 2017. 
	In 2021, he obtained a Ph.D. in mechanical engineering, also from the University of Stuttgart, Germany.
	He is currently a postdoctoral researcher at the Institute for Dynamic Systems and Control (IDSC) at ETH Zürich.
	His research interests are in the area of model predictive control and control and estimation for nonlinear uncertain systems. 
\end{IEEEbiography}

\begin{IEEEbiography}
	[{\includegraphics[width=1in,height=1.25in,clip,keepaspectratio]{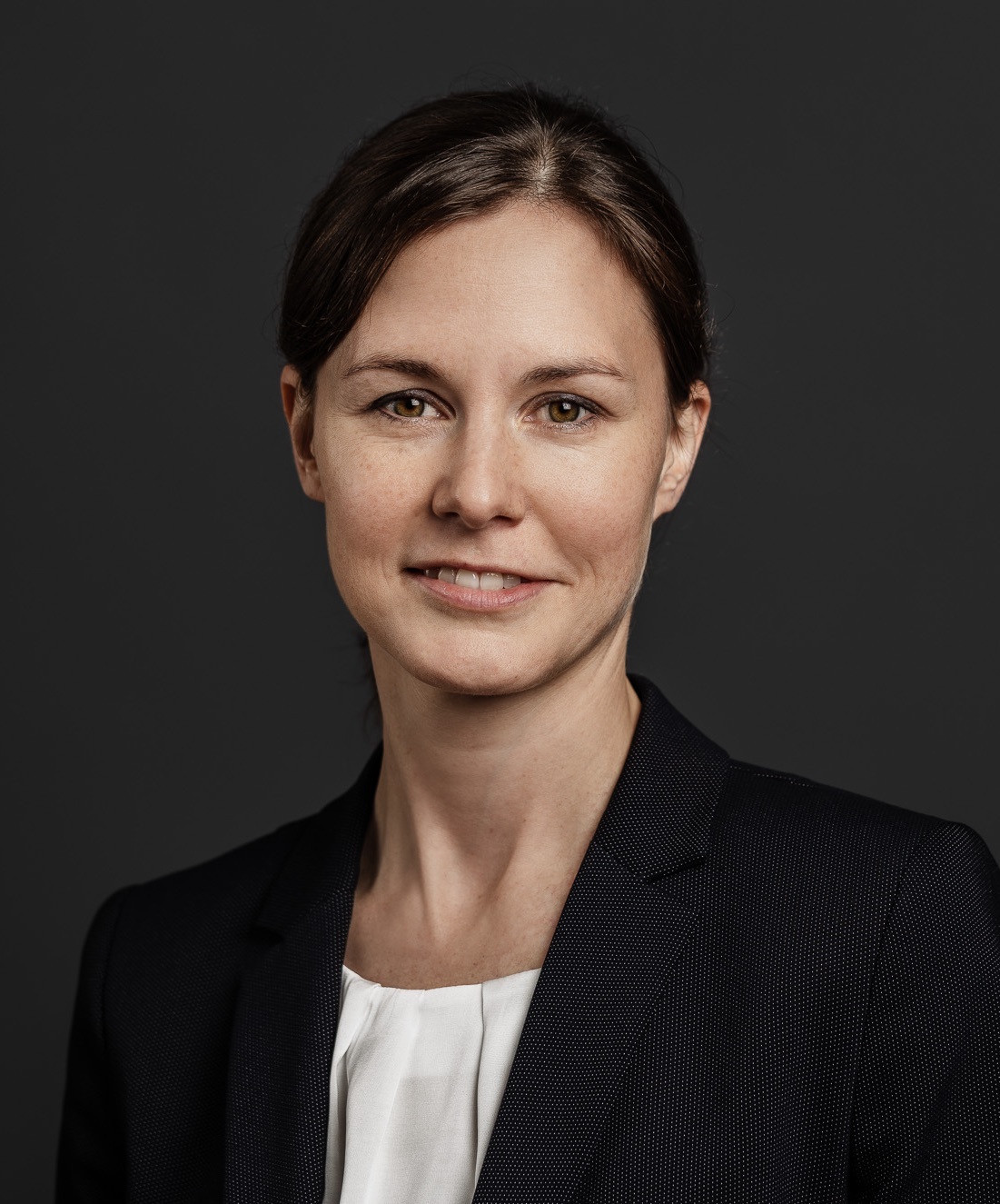}}]{Melanie N. Zeilinger}
 	is an Assistant Professor at ETH Zürich, Switzerland. 
	She received the Diploma degree in engineering cybernetics from the University of Stuttgart, Germany, in 2006, and the Ph.D. degree with honors in electrical engineering from ETH Zürich, Switzerland, in 2011. 
	From 2011 to 2012 she was a Postdoctoral Fellow with the Ecole Polytechnique Federale de Lausanne (EPFL), Switzerland.
	She was a Marie Curie Fellow and Postdoctoral Researcher with the Max Planck Institute for Intelligent 	Systems, Tübingen, Germany until 2015 and with the Department of Electrical Engineering and Computer Sciences at the University of California at Berkeley, CA, USA, from 2012 to 2014. 
	From 2018 to 2019 she was a professor at the University of Freiburg, Germany. 
	Her current research interests include safe learning-based control, as well as distributed control and optimization, with applications to robotics and human-in-the loop control.
\end{IEEEbiography}

\begin{IEEEbiography}
	[{\includegraphics[width=1in,height=1.25in,clip,keepaspectratio]{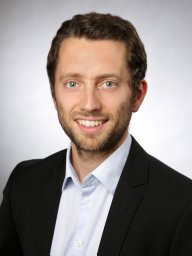}}]{Matthias A. M\"uller}
	received a Diploma degree in Engineering Cybernetics from the University of Stuttgart, Germany, and an M.S. in Electrical and Computer Engineering from the University of Illinois at Urbana-Champaign, US, both in 2009.
	In 2014, he obtained a Ph.D. in Mechanical Engineering, also from the University of Stuttgart, Germany, for which he received the 2015 European Ph.D. award on control for complex and heterogeneous systems.
	Since 2019, he is director of the Institute of Automatic Control and full professor at the Leibniz University Hannover, Germany. 
	He obtained an ERC Starting Grant in 2020 and is recipient of the inaugural Brockett-Willems Outstanding Paper Award for the best paper published in Systems \& Control Letters in the period 2014-2018. His research interests include nonlinear control and estimation, model predictive control, and data-/learning-based control, with application in different fields including biomedical engineering.
\end{IEEEbiography}
\appendix
\subsection{Technical details of Theorem~\ref{thm:dIOSS_Lyap}}
\label{app:dIOSS_proof_thm}

In the following, we provide further technical details of Theorem~\ref{thm:dIOSS_Lyap}, including the proof itself and the modifications required by Remark~\ref{rem:extensions}.

\begin{proof}[Proof of Theorem~\ref{thm:dIOSS_Lyap}]
	The proof consists of three parts. First, we establish the dissipation inequality \eqref{eq:dIOSS_quad_gains} and then derive the bounds \eqref{eq:dIOSS_quad_bounds}, where we initially assume that the conditions \eqref{eq:dIOSS_LMI_1}-\eqref{eq:dIOSS_P1P2} hold globally on $\mathbb{R}^n\times\mathbb{R}^m\times\mathbb{R}^q$.
	Finally, we show that the corresponding results also hold if the conditions are enforced on the subset $\mathbb{X}\times\mathbb{U}\times\mathbb{W}$ only.

	\textbf{Part I:} Consider two arbitrary points $(x,u,w,y)$ and $(\tilde{x},u,\tilde{w},\tilde{y})$ each of which is an element of $\mathbb{R}^n\times\mathbb{R}^m\times\mathbb{R}^q\times\mathbb{R}^p$.
	Define a smooth path $c : [0,1] \rightarrow \mathbb{R}^n$ parametrized by $s$ joining $x$ to $\tilde{x}$ with $c(0) = x$ and $c(1) = \tilde{x}$.
	Define the smooth path of disturbances $\omega(s)$ joining $\omega(0) = w$ and $\omega(1) = \tilde{w}$ by the straight line
	\begin{equation}\label{eq:dIOSS_omega}
		\omega(s) = w + s(\tilde{w}-w), \quad s\in[0,1].
	\end{equation}
	Note that this particular choice is valid since the disturbance can generally be treated as an external variable that does not depend on any dynamics; therefore, the path connecting $w$ and $\tilde{w}$ can be of arbitrary form.	
	Given the tuple $(c(s),u,\omega(s))$, we can apply the dynamics \eqref{eq:sys_w} and obtain
	\begin{subequations}\label{eq:dIOSS_path_dyn}
		\begin{align}
			c^+(s) &= f(c(s),u,\omega(s))\label{eq:dIOSS_path_dyn_1},\\
			\zeta(s) &= h(c(s),u,\omega(s))\label{eq:dIOSS_path_dyn_2},
		\end{align}
	\end{subequations}
	where the corresponding output $\zeta$ yields a smooth path joining $\zeta(0)=y$ and $\zeta(1) = \tilde{y}$.
	By differentiating \eqref{eq:dIOSS_path_dyn} with respect to $s\in[0,1]$, from the chain rule and the linearizations \eqref{eq:dIOSS_ABCD} we obtain the differential dynamics
	\begin{subequations}\label{eq:diff_dyn}
		\begin{align}
		\delta_x^+ 
		&= A(c(s),u,\omega(s))\delta_x 
		+ B(c(s),u,\omega(s))\delta_w, \label{eq:diff_dyn_1}\\
		\delta_y 
		&= C(c(s),u,\omega(s))\delta_x 
		+ D(c(s),u,\omega(s))\delta_w, \label{eq:diff_dyn_2}
		\end{align}
	\end{subequations}
	where the path derivatives are defined as $\delta_x^+:=dc^+/ds(s)$, $\delta_x:=dc/ds(s)$, $\delta_w:=d\omega/ds(s)$, and $\delta_y:=d\zeta/ds(s)$.
	Formally, each $\delta_i$ with $i \in \lbrace x,w,y\rbrace$ denotes a vector on the tangent space of the domain of $i$ at $i$, cf. \cite{manchester2017control,manchester2018robustCCM}.
	We make the following claim.
	\begin{claim}\label{claim:dIOSS_1}
		Let \eqref{eq:dIOSS_LMI_1} hold for some $(x,u,w)\in\mathbb{R}^n\times\mathbb{R}^m\times\mathbb{R}^q$. Then, $V(x,\delta_x) = \delta_x^\top P(x) \delta_x$ satisfies
		\begin{equation}\label{eq:dIOSS_dissip}
			V(x^+,\delta_x^+) \leq  \eta V(x,\delta_x) + \|\delta_w \|^2_Q + \|\delta_y \|^2_R.
		\end{equation}
	\end{claim}
	\begin{proof}
		By applying the definition of $V$ together with the differential dynamics (\ref{eq:diff_dyn_1})-(\ref{eq:diff_dyn_2}) to \eqref{eq:dIOSS_dissip}, we obtain
		\begin{align*}
			&\begin{pmatrix} \delta_x \\ \delta_w\end{pmatrix}^\top
			\begin{pmatrix} A^\top P_+A-\eta P & A^\top P_+B	\\ B^\top P_+A & B^\top P_+ B\end{pmatrix}
			\begin{pmatrix} \delta_x \\ \delta_w \end{pmatrix} \\
			& \quad \preceq
			\begin{pmatrix} \delta_x \\ \delta_w \end{pmatrix}^\top
			\begin{pmatrix} C^\top RC & C^\top RD \\  D^\top RC & Q+D^\top RD \end{pmatrix}
			\begin{pmatrix} \delta_x \\ \delta_w \end{pmatrix}, 
		\end{align*}
		which clearly is equivalent to \eqref{eq:dIOSS_LMI_1}.
	\end{proof}

	Consequently, by definition of the differential storage function $V$ and the path derivatives $(\delta_x,\delta_w,\delta_y)$, from \eqref{eq:dIOSS_dissip} it follows that
	\begin{equation}\label{eq:dIOSS_diff_c}
		\left\|\frac{dc^+}{ds}(s)\right\|_{P_+}^2
		\leq \eta\left\|\frac{dc}{ds}(s)\right\|_P^2
		+ \left\|\frac{d\omega}{ds}(s)\right\|_Q^2
		+ \left\|\frac{d\zeta}{ds}(s)\right\|_R^2.
	\end{equation}
	This differential property can now be transformed into an incremental property by integration over $s\in[0,1]$ and utilizing tools from Riemannian geometry.
	In particular, we treat $P$ as a Riemannian\footnote{
		A Riemannian metric $P:\mathbb{R}^n\rightarrow\mathbb{R}^{n\times n}$ is a symmetric covariant 2-tensor with positive definite values that defines local notions of length, angle, and orthogonality by the inner product $\langle \delta_1,\delta_2\rangle_x = \delta_1^\top P(x) \delta_2$ for any two tangent vectors $\delta_1,\delta_2$, cf.  \cite{sanfelice2012metric,manchester2017control}, and compare also \cite{gallot2004riemannian} for further details.		 
	} metric with which the manifold $\mathbb{R}^n$ is endowed.
	Let 
	\begin{equation}\label{eq:dIOSS_energy}
		E(c) := \int_0^1 \frac{dc}{ds}(s)^\top P(c(s)) \frac{dc}{ds}(s) ds
	\end{equation}
	denote the Riemannian energy associated with the path $c$.
	The minimizer of $E(c)$ over all possible smooth paths joining $c(0)$ to $c(1)$ is given by a (maybe non-unique) \textit{geodesic} $\gamma$, existence of which is ensured by the uniformly boundedness of $P$ in~\eqref{eq:dIOSS_P1P2}, cf. \cite[Lem. 1]{manchester2017control}, compare also \cite[Lem. A.1]{sanfelice2012metric}.
	
	\addtolength{\textheight}{-3.333cm} 
	
	Now, consider \eqref{eq:dIOSS_diff_c} and choose $c = \gamma$; it hence follows that $c^+(s) = f(\gamma(s),u,\omega(s))$ by~\eqref{eq:dIOSS_path_dyn_1}.
	Let $\gamma^+$ denote the geodesic at the subsequent time instance joining the successor states $x^+$ and $\tilde{x}^+$, where we point out that in general $c^+\neq\gamma^+$.
	However, by integration of \eqref{eq:dIOSS_diff_c} over $s\in[0,1]$ and the definition of the Riemannian energy \eqref{eq:dIOSS_energy}, we obtain	
	\begin{align}
		E(\gamma^+)
		&\leq E(c^+) \notag \\ &\leq \eta E(\gamma) 
		+ \int_0^1 \left\|\frac{d\omega}{ds}(s)\right\|_Q^2 ds
		+ \int_0^1 \left\|\frac{d\zeta}{ds}(s)\right\|_R^2 ds, \label{eq:dIOSS_int}
	\end{align}
	where the first inequality used the fact that $c^+$ is feasible candidate curve providing an upper bound for the (minimal) energy $E(\gamma^+)$.
	In the following, we show that $W_\delta(x,\hat{x})=E(\gamma)$ satisfies the dissipation inequality \eqref{eq:dIOSS_quad_gains}.
	First, exploiting our particular choice of $\omega$ in \eqref{eq:dIOSS_omega} yields
	\begin{equation}
		\int_0^1 \left\|\frac{d\omega}{ds}(s)\right\|_Q^2 ds 
		= \int_{0}^{1}\left\|\tilde{w}-w\right\|^2_Qds = \left\|\tilde{w}-w\right\|^2_Q. \label{eq:dIOSS_int_w}
	\end{equation}
	Now we focus on the output term in \eqref{eq:dIOSS_int} and make the following claim.
	\begin{claim}\label{claim:dIOSS_2}
		The derivative $d\zeta/ds(s)$ is constant in $s\in[0,1]$.
	\end{claim}
	\begin{proof}		
		Given $\gamma$ and $\phi$, we can define the geodesic in transformed coordinates $\bar{\gamma} := \phi(\gamma)$.
		From~\eqref{eq:dIOSS_path_dyn_2}, we have that
		\begin{equation}\label{eq:dIOSS_bar_gamma_2}
			\zeta(s) = h(\phi^{-1}(\bar{\gamma}(s)),u,\omega(s)) = \bar{h}(\bar{\gamma}(s),u,\omega(s))
		\end{equation}
		for all $s\in[0,1]$.
		Taking the derivative of \eqref{eq:dIOSS_bar_gamma_2} with respect to $s\in[0,1]$ using the chain rule yields
		\begin{align*}
			&\frac{d\zeta}{ds}(s)
			= \frac{d\bar{h}}{ds}(\bar{\gamma}(s),u,\omega(s)) \notag \\
			& \ \ = \frac{\partial \bar{h}}{\partial \bar{x}}(\bar{\gamma}(s),u,\omega(s))\frac{d\bar{\gamma}}{ds}(s) + \frac{\partial \bar{h}}{\partial w}(\bar{\gamma}(s),u,\omega(s))\frac{d\omega}{ds}(s).
		\end{align*}
		Assumption~\ref{ass:dIOSS} ensures that $\bar{h}$ is affine in $\bar{x},w$, and consequently, the partial derivatives $\partial \bar{h}/\partial \bar{x}(\bar{\gamma}(s),u,\omega(s))$ and $\partial \bar{h}/\partial w(\bar{\gamma}(s),u,\omega(s))$ do not depend on $s$; furthermore, $\partial \bar{h}/\partial \bar{x}_i(\bar{\gamma}(s),u,\omega(s))=0$ for all $i=1,\ldots,n-p$.
		Since, in addition, $d\omega/ds(s)$ is constant in $s\in[0,1]$ due to ~\eqref{eq:dIOSS_omega}, it remains to show that this is also the case for $d\bar{\gamma}_i/ds(s)$ for all $i=n-p+1,\ldots,n$.
		
		To this end, recall that $\bar{\gamma}=\phi(\gamma)$. Hence, by the chain rule,
		\begin{equation*}
			\frac{d\bar\gamma}{ds}(s) = \frac{\partial \phi}{\partial x} (\gamma(s))\frac{d\gamma}{ds}(s).
		\end{equation*}
		Due to our choice of $P$ in \eqref{eq:dIOSS_Px}, it therefore holds that
		\begin{align*}
			\bar{E}&(\bar\gamma) 
			:= \int_0^1 \frac{d\bar\gamma}{ds}(s)^\top \bar{P}(\bar\gamma(s)) \frac{d\bar\gamma}{ds}(s) ds\\
			& = \int_0^1 \frac{d\gamma}{ds}(s)^\top\frac{\partial \phi}{\partial x}(\gamma(s))^\top
			\bar{P}(\phi(\gamma(s))) 
			\frac{\partial \phi}{\partial x}(\gamma(s))\frac{d\gamma}{ds}(s)ds\\
			& = \int_0^1 \frac{d\gamma}{ds}(s)^\top P(\gamma(s))\frac{d\gamma}{ds}(s) ds = E(\gamma).
		\end{align*}
		Thus, given a minimizing geodesic $\gamma$ for $E(\gamma)$, the curve~$\bar{\gamma}$ is a minimizing geodesic for $\bar{E}(\bar{\gamma})$ (by contradiction).		
		Consequently, we have that $\bar{\gamma}(s)=\phi(\gamma(s))$ is a solution to the geodesic equation \cite[Def. 2.77]{gallot2004riemannian}, i.e., to the differential system
		\begin{align}\label{eq:dIOSS_geodesic}
			\frac{d^2\bar{\gamma}_k}{ds^2}(s) - \sum_{i,j}\bar{\Gamma}_{i,j}^k(\bar{\gamma}(s))\frac{d\bar{\gamma}_i}{ds}(s)\frac{d\bar{\gamma}_j}{ds}(s) = 0, \quad k = 1,\ldots,n.
		\end{align}
		The objects $\bar{\Gamma}_{i,j}^k$ represent the Christoffel symbols associated with the metric $\bar{P}$ which are, following \cite[Prop 2.54]{gallot2004riemannian} and \cite[App. A1.1]{sanfelice2021metric}, defined by
		\begin{equation}\label{eq:dIOSS_christoffel}
			\bar{\Gamma}_{i,j}^k(\bar{x}) = \frac{1}{2}\sum_{a=1}^{n}\bar{Y}_{k,a}(\bar{x})
			\left(\frac{\partial \bar{P}_{a,i}}{\partial \bar{x}_j}(\bar{x})
			{+}\frac{\partial \bar{P}_{a,j}}{\partial \bar{x}_i}(\bar{x})
			-\frac{\partial \bar{P}_{i,j}}{\partial \bar{x}_a}(\bar{x})\right)
		\end{equation}
		with the shorthand notation $\bar{Y}(\bar{x})=\bar{P}(\bar{x})^{-1}$ and $\bar{Y}_{k,a}$ the $(k,a)$-element of $\bar{Y}$.
		Note that in \eqref{eq:dIOSS_geodesic}, we are only interested in the states of the geodesic $\bar{\gamma}$ that appear in the output \eqref{eq:dIOSS_bar_gamma_2}, i.e., $\bar{\gamma}_k$ for all $k=n-p+1,\ldots,n$. For ease of notation, let us define $r:=n-p+1$ for the remainder of this proof.
		Calculating the respective Christoffel symbols reveals that
		\begin{equation}\label{eq:dIOSS_christoffel_r}
			\bar{\Gamma}_{ij}^k=0, \quad k=r,\ldots,n,
		\end{equation} 
		which is a direct consequence of the proposed block-diagonal structure of $\bar{P}$ in \eqref{eq:dIOSS_Pblock}; to see this, note the following:
		First, the fact that $\bar{P}$ is block-diagonal implies that also $\bar{Y}=\bar{P}^{-1}$ is block-diagonal, and thus $\bar{Y}_{i,j} = \bar{P}_{i,j} = 0$ for $i<r$ and $j\geq r$ (and vice versa); second, each derivative $\partial \bar{P}_{i,j}/\partial\bar{x}_a = 0$ if $a\geq r$ since $\bar{P}$ is independent of $\bar{x}_y$; third, each derivative $\partial \bar{P}_{a,i}/\partial\bar{x}_j = 0$ if $a,i\geq r,$ and $j<r$ since $\bar{P}_y$ is constant.
		
		Consequently, from \eqref{eq:dIOSS_christoffel_r}, we have that all the Christoffel symbols affecting the states $\bar{\gamma}_i$, $i=r,\ldots,n$ vanish, and hence our special choice of $\bar{P}$ leads to a decoupling of the geodesic equation~\eqref{eq:dIOSS_geodesic}; in particular, we obtain the simple second-order homogeneous differential equation
		\begin{equation*}
			\frac{d^2\bar{\gamma}_k}{ds^2}(s) = 0, \quad k= r,\ldots,n,
		\end{equation*}
		which directly implies that $d\bar{\gamma}_i/ds(s)$ is constant in $s\in[0,1]$ for all $i=r,\ldots,n$ and hence yields the desired result.
	\end{proof}
	
	Consequently, the output term in \eqref{eq:dIOSS_int} consists only of terms constant in $s\in[0,1]$.
	Hence, by the Fundamental Theorem of Calculus, we obtain
	\begin{align}
		\int_0^1 \left\|\frac{d\zeta}{ds}(s)\right\|_R^2 ds
		&= (\zeta(1)-\zeta(0))^\top R (\zeta(1)-\zeta(0)) \nonumber \\
		&= \left\|\tilde{y}-y\right\|^2_R \label{eq:dIOSS_int_y}.
	\end{align}
	Applying (\ref{eq:dIOSS_int_w}) and (\ref{eq:dIOSS_int_y}) to (\ref{eq:dIOSS_int}) then yields
	\begin{equation}\label{eq:dIOSS_dissip_E}
		E(\gamma^+)
		\leq \eta E(\gamma)
		+ \|w-\tilde{w}\|_Q^2
		+ \|y-\tilde{y}\|_R^2,
	\end{equation}
	which establishes the dissipation inequality \eqref{eq:dIOSS_quad_gains} with $W_\delta(x,\hat{x})=E(\gamma)$.
	
	\textbf{Part II:} We now show satisfaction of \eqref{eq:dIOSS_quad_bounds} and start with the upper bound.
	Note that since $\gamma$ is the path of minimum energy joining $x$ to $\tilde{x}$, every other path yields a higher amount of energy, which clearly also applies to the straight line $l(s) = x+s(\tilde{x}-x)$. Therefore,
	\begin{equation}
		E(\gamma) \leq E(l) = \int_0^1 (x-\tilde{x})^\top P(l(s)) (x-\tilde{x}) ds
		\leq \|x-\tilde{x}\|^2_{P_2}, \label{eq:dIOSS_P2}
	\end{equation}
	where the last step follows from uniform boundedness of $P$~\eqref{eq:dIOSS_P1P2}.
	For the lower bound, again by uniform boundedness of $P$, we have
	\begin{align}
		E(\gamma) &\geq \int_0^1  \frac{\partial \gamma}{ds}(s)^\top P_1 \frac{\partial \gamma}{ds}(s) ds \label{eq:dIOSS_P1_intermediate} \\
		&\geq \int_0^1 \frac{\partial l}{ds}(s)^\top P_1 \frac{\partial l}{ds}(s) ds = \|x-\tilde{x}\|_{P_1}^2, \label{eq:dIOSS_P1}
	\end{align}
	where for the second inequality we exploited the fact that the minimizer of the expression on the right hand side of \eqref{eq:dIOSS_P1_intermediate} is given by the straight line $l$ since $P_1$ is constant.
	To verify this, recall that each minimizer of the Riemannian energy $E$ solves the geodesic equation \eqref{eq:dIOSS_geodesic}; now observe that all the Christoffel symbols \eqref{eq:dIOSS_christoffel} vanish if the underlying metric is constant.
	Therefore, \eqref{eq:dIOSS_P2} and \eqref{eq:dIOSS_P1} establish \eqref{eq:dIOSS_quad_bounds}.
	Together with Part I, we can thus conclude that $W(x,\tilde{x})=E(\gamma)$ is a $\delta$-IOSS Lyapunov function satisfying \eqref{eq:dIOSS_quad_bounds} and \eqref{eq:dIOSS_quad_gains} for all $(x,u,w,y),(\tilde{x},u,\tilde{w},\tilde{y}) \in \mathbb{R}^n\times\mathbb{R}^m\times\mathbb{R}^q\times\mathbb{R}^p$.
	
	\textbf{Part III:}
	Finally, we note that the results from Part~I and Part~II, (including the Claims \ref{claim:dIOSS_1} and \ref{claim:dIOSS_2}) can be easily restricted to any subset $\mathbb{X}\times\mathbb{U}\times\mathbb{W}\times\mathbb{Y}$ if it is ensured that the minimizing geodesic connecting any two points on each of the subsets $\mathbb{X}$ and $\mathbb{W}$ stays in the respective subset for all $s \in [0,1]$.	 
	This is indeed the case for $\mathbb{X}$ being weakly geodesically convex (cf.~\cite[Def. 2.6]{sanfelice2012metric}) and $\mathbb{W}$ being convex (as long as $\omega$ is chosen according to \eqref{eq:dIOSS_omega}).
	Provided that this applies, if conditions \eqref{eq:dIOSS_LMI_1}-\eqref{eq:dIOSS_P1P2} are enforced on the subset $\mathbb{X}\times\mathbb{U}\times\mathbb{W}$, we have that $W_\delta(x,\tilde{x}) = E(\gamma)$ is a quadratically bounded $\delta$-IOSS Lyapunov function satisfying Assumption~\ref{ass:IOSS_quadratically_bounded} for all $(x,u,w,y),(\tilde{x},u,\tilde{w},\tilde{y}) \in \mathbb{X}\times\mathbb{U}\times\mathbb{W}\times\mathbb{Y}$, which completes this proof.	
\end{proof}

We now discuss the modifications of the proof of Theorem~\ref{thm:dIOSS_Lyap} that are necessary to allow for the slightly more general metric~$\bar{P}$ from Remark~\ref{rem:extensions}.

\begin{remark}\label{rem:extensions_details}
	If $\bar{P}$ is chosen according to \eqref{eq:dIOSS_P_extension}, i.e.,
	\begin{equation*}
		\bar{P}(\bar{x}) = \begin{pmatrix} \bar{P}_x(\bar{x}_x) & 0\\ 0 & \bar{P}_y(\bar{x}_y)\end{pmatrix},
	\end{equation*}
	we have that the geodesic $\bar{\gamma}$ minimizes the two independent functionals
	\begin{align}\label{eq:dIOSS_rem}
	\bar{E}(\bar{\gamma}) &= 
	\int_0^1 \frac{d\bar\gamma_x}{ds}(s)^\top \bar{P}_x(\bar\gamma_x(s)) \frac{d\bar\gamma_x}{ds}(s) ds \\ \nonumber
	& \qquad + \int_0^1 \frac{d\bar\gamma_y}{ds}(s)^\top \bar{P}_y(\bar\gamma_y(s)) \frac{d\bar\gamma_y}{ds}(s) ds.
	\end{align}
	Note that a direct consequence of $\bar{P}_y(\bar{x}_y)$ not being constant is that Claim~\ref{claim:dIOSS_2} does not hold in this case.
	However, since $R \preceq \bar{P}_y(\bar x_y)$ by Remark~\ref{rem:extensions}, the output functional in \eqref{eq:dIOSS_diff_c} can be bounded by
	\begin{align*}
	&\int_0^1 \left\|\frac{d\zeta}{ds}(s)\right\|_{R}^2 ds
	\leq 
	\int_0^1 \frac{d\bar\gamma_y}{ds}(s)^\top \bar{P}_y(\bar\gamma_y(s)) \frac{d\bar\gamma_y}{ds}(s) ds,
	\end{align*}
	i.e., the same functional that also appears in \eqref{eq:dIOSS_rem} and hence is minimized by $\bar{\gamma}$.
	Then, by following similar arguments as in the second part of the proof of Theorem~\ref{thm:dIOSS_Lyap} (in particular, exploiting uniform boundedness of $\bar{P}_{y}$ according to Remark~\ref{rem:extensions}), one can show that $\int_0^1 \left\|d\zeta/ds(s)\right\|_{R}^2 ds \leq \left\|\tilde{y}-y\right\|_{\bar{P}_{y,2}}^2$ and subsequently derive a similar $\delta$-IOSS Lyapunov function as in Theorem~\ref{thm:dIOSS_Lyap} that satisfies Assumption~\ref{ass:IOSS_quadratically_bounded}.
\end{remark}

\subsection{Proof of Corollary \ref{cor:dIOSS_Lyap}}
\label{app:dIOSS_proof_cor}

\begin{proof}
	The result follows immediately by setting $\bar{x}_y=\bar{x}=\phi(x)=x$ in the proof of Theorem~\ref{thm:dIOSS_Lyap}.
	As a direct consequence, we obtain \eqref{eq:dIOSS_dissip_E} with $E(\gamma)=\|x-\tilde{x}\|_P^2$ and $E(\gamma^+)=\|x^+-\tilde{x}^+\|_P^2$ since $P$ is constant (resulting in the geodesics being straight lines), which lets us conclude that $W_\delta(x,\tilde{x})=\|x-\tilde{x}\|_P^2$ is a quadratic \mbox{$\delta$-IOSS} Lyapunov function that satisfies Assumption~\ref{ass:IOSS_quadratically_bounded} with $P_1=P_2=P$ for all  $(x,u,w,y),(\tilde{x},u,\tilde{w},\tilde{y})\in\mathbb{X}\times\mathbb{U}\times\mathbb{W}\times\mathbb{Y}$.
\end{proof}


\begin{thebibliography}{10}
\providecommand{\url}[1]{#1}
\csname url@samestyle\endcsname
\providecommand{\newblock}{\relax}
\providecommand{\bibinfo}[2]{#2}
\providecommand{\BIBentrySTDinterwordspacing}{\spaceskip=0pt\relax}
\providecommand{\BIBentryALTinterwordstretchfactor}{4}
\providecommand{\BIBentryALTinterwordspacing}{\spaceskip=\fontdimen2\font plus
\BIBentryALTinterwordstretchfactor\fontdimen3\font minus
  \fontdimen4\font\relax}
\providecommand{\BIBforeignlanguage}[2]{{%
\expandafter\ifx\csname l@#1\endcsname\relax
\typeout{** WARNING: IEEEtran.bst: No hyphenation pattern has been}%
\typeout{** loaded for the language `#1'. Using the pattern for}%
\typeout{** the default language instead.}%
\else
\language=\csname l@#1\endcsname
\fi
#2}}
\providecommand{\BIBdecl}{\relax}
\BIBdecl

\bibitem{manchester2021observer}
B.~Yi, R.~Wang, and I.~R. Manchester, ``Reduced-order nonlinear observers via
  contraction analysis and convex optimization,'' \emph{{IEEE} Trans. Automat.
  Contr.}, vol.~67, no.~8, pp. 4045--4060, 2022.

\bibitem{sanfelice2021metric}
R.~G. {Sanfelice} and L.~{Praly}, ``Convergence of nonlinear observers on
  $\mathbb{R}^{n}$ with a {Riemannian} metric ({Part III}),'' \emph{arXiv
  preprint arXiv:2102.08340}, 2021.

\bibitem{astolfi2022constrained}
D.~Astolfi, P.~Bernard, R.~Postoyan, and L.~Marconi, ``Constrained state
  estimation for nonlinear systems: A redesign approach based on convexity,''
  \emph{{IEEE} Trans. Automat. Contr.}, 2022.

\bibitem{allan2019moving}
D.~A. Allan and J.~B. Rawlings, ``Moving horizon estimation,'' in
  \emph{Handbook of Model Predictive Control}, S.~V. Raković and W.~S. Levine,
  Eds.\hskip 1em plus 0.5em minus 0.4em\relax Basel, Switzerland: Birkhäuser,
  2019, pp. 99--124.

\bibitem{rawlings2020model}
J.~B. Rawlings, D.~Q. Mayne, and M.~Diehl, \emph{Model Predictive Control:
  Theory, Computation, and Design}, 2nd~ed.\hskip 1em plus 0.5em minus
  0.4em\relax Santa Barbara, CA, USA: Nob Hill Publish., LLC, 2020, 3rd
  printing.

\bibitem{michalska1995moving}
H.~Michalska and D.~Q. Mayne, ``Moving horizon observers and observer-based
  control,'' \emph{{IEEE} Trans.\ Automat.\ Contr}, vol.~40, pp. 995--1006,
  1995.

\bibitem{alessandri2008moving}
A.~Alessandri, M.~Baglietto, and G.~Battistelli, ``Moving-horizon state
  estimation for nonlinear discrete-time systems: {N}ew stability results and
  approximation schemes,'' \emph{Automatica}, vol.~44, no.~7, pp. 1753--1765,
  2008.

\bibitem{rao2003constrained}
C.~V. Rao, J.~B. Rawlings, and D.~Q. Mayne, ``Constrained state estimation for
  nonlinear discrete-time systems: Stability and moving horizon
  approximations,'' \emph{IEEE Trans.\ Automat.\ Contr}, vol.~48, no.~2, pp.
  246--258, 2003.

\bibitem{gharbi2021proximity}
M.~Gharbi, F.~Bayer, and C.~Ebenbauer, ``Proximity moving horizon estimation
  for discrete-time nonlinear systems,'' \emph{IEEE Contr. Syst. Lett.},
  vol.~5, no.~6, pp. 2090--2095, 2021.

\bibitem{schiller2021suboptimal}
J.~D. Schiller and M.~A. M\"{u}ller, ``Suboptimal nonlinear moving horizon
  estimation,'' \emph{{IEEE} Trans.\ Automat.\ Contr.}, vol.~68, no.~4, pp.
  2199--2214, 2023.

\bibitem{sontag1997output}
E.~D. Sontag and Y.~Wang, ``Output-to-state stability and detectability of
  nonlinear systems,'' \emph{Syst. Contr. Lett.}, vol.~29, no.~5, pp. 279--290,
  1997.

\bibitem{rawlings2012optimization}
J.~B. Rawlings and L.~Ji, ``Optimization-based state estimation: Current status
  and some new results,'' \emph{J.\ Proc.\ Contr.}, vol.~22, pp. 1439--1444,
  2012.

\bibitem{muller2016nonlinear}
M.~A. M{\"u}ller, ``Nonlinear moving horizon estimation for systems with
  bounded disturbances,'' in \emph{Proc. Amer. Contr. Conf. (ACC)}, 2016, pp.
  883--888.

\bibitem{hu2017robust}
W.~Hu, ``Robust stability of optimization-based state estimation,'' \emph{arXiv
  preprint arXiv:1702.01903v3}, 2017.

\bibitem{muller2017nonlinear}
M.~A. M{\"u}ller, ``Nonlinear moving horizon estimation in the presence of
  bounded disturbances,'' \emph{Automatica}, vol.~79, pp. 306--314, 2017.

\bibitem{allan2019lyapunov}
D.~A. Allan and J.~B. Rawlings, ``A {L}yapunov-like function for full
  information estimation,'' in \emph{Proc. Amer. Contr. Conf. (ACC)}, 2019, pp.
  4497--4502.

\bibitem{allan2021FIE}
------, ``Robust stability of full information estimation,'' \emph{{SIAM} J.
  Contr. Optim.}, vol.~59, no.~5, pp. 3472--3497, 2021.

\bibitem{knuefer2021MHE}
S.~Kn\"ufer and M.~A. M\"uller, ``Nonlinear full information and moving horizon
  estimation: Robust global asymptotic stability,'' \emph{Automatica}, vol.
  150, p. 110603, 2023.

\bibitem{hu2021MHE}
W.~Hu, ``Generic stability implication from full information estimation to
  moving-horizon estimation,'' \emph{{IEEE} Trans. Automat. Contr.}, 2023,
  early access, doi: 10.1109/TAC.2023.3277315.

\bibitem{knufer2018robust}
S.~Kn{\"u}fer and M.~A. M{\"u}ller, ``Robust global exponential stability for
  moving horizon estimation,'' in \emph{Proc.\ 57th IEEE Conf.\ Decis. Contr.
  (CDC)}, 2018, pp. 3477--3482.

\bibitem{koehler2021Output}
J.~K\"{o}hler, M.~A. M{\"u}ller, and F.~Allg{\"o}wer, ``Robust output feedback
  model predictive control using online estimation bounds,'' \emph{arXiv
  preprint arXiv:2105.03427}, 2021.

\bibitem{allan2020lyapunov}
D.~A. Allan, ``A {L}yapunov-like function for analysis of model predictive
  control and moving horizon estimation,'' Ph.D. dissertation, Univ.
  Wisconsin-Madison, 2020.

\bibitem{allan2021detect}
D.~A. Allan, J.~B. Rawlings, and A.~R. Teel, ``Nonlinear detectability and
  incremental input/output-to-state stability,'' \emph{{SIAM} J. Contr.
  Optim.}, vol.~59, no.~4, pp. 3017--3039, 2021.

\bibitem{knuefer2020time}
S.~Kn\"ufer and M.~A. M\"uller, ``Time-discounted incremental
  input/output-to-state stability,'' in \emph{Proc. 59th IEEE Conf. Decis.
  Contr. (CDC)}, 2020, pp. 5394--5400.

\bibitem{tenny2002efficient}
M.~J. Tenny and J.~B. Rawlings, ``Efficient moving horizon estimation and
  nonlinear model predictive control,'' in \emph{Proc. Amer. Contr. Conf.
  (ACC)}, 2002.

\bibitem{angeli2002lyapunov}
D.~Angeli, ``A {L}yapunov approach to incremental stability properties,''
  \emph{IEEE Trans.\ Automat.\ Contr.}, vol.~47, pp. 410--421, 2002.

\bibitem{ahmadi2008non}
A.~A. Ahmadi and P.~A. Parrilo, ``Non-monotonic {L}yapunov functions for
  stability of discrete time nonlinear and switched systems,'' in \emph{Proc.\
  47th IEEE Conf.\ Decis. Contr. (CDC)}, 2008, pp. 614--621.

\bibitem{Grimm2005model}
G.~Grimm, M.~Messina, S.~Tuna, and A.~Teel, ``Model predictive control: for
  want of a local control {L}yapunov function, all is not lost,'' \emph{IEEE
  Trans. Automat. Contr}, vol.~50, no.~5, pp. 546--558, 2005.

\bibitem{koehler2021dynamic}
J.~K\"{o}hler, ``Analysis and design of {MPC} frameworks for dynamic operation
  of nonlinear constrained systems,'' Ph.D. dissertation, Universit{\"a}t
  Stuttgart, 2021.

\bibitem{sontag1989smooth}
E.~Sontag, ``Smooth stabilization implies coprime factorization,'' \emph{{IEEE}
  Trans.\ Automat.\ Contr.}, vol.~34, no.~4, pp. 435--443, 1989.

\bibitem{boyd1985fading}
S.~Boyd and L.~Chua, ``Fading memory and the problem of approximating nonlinear
  operators with volterra series,'' \emph{{IEEE} Trans. Circuits Syst.},
  vol.~32, no.~11, pp. 1150--1161, 1985.

\bibitem{maass2000neural}
W.~Maass and E.~D. Sontag, ``Neural systems as nonlinear filters,''
  \emph{Neural Comput.}, vol.~12, no.~8, pp. 1743--1772, 2000.

\bibitem{sorenson1971recursive}
H.~Sorenson and J.~Sacks, ``Recursive fading memory filtering,'' \emph{Inf.
  Sci.}, vol.~3, no.~2, pp. 101--119, 1971.

\bibitem{lohmiller1998contraction}
W.~Lohmiller and J.-J.~E. Slotine, ``On contraction analysis for non-linear
  systems,'' \emph{Automatica}, vol.~34, pp. 683--696, 1998.

\bibitem{manchester2014transverse}
I.~R. Manchester and J.-J.~E. Slotine, ``Transverse contraction criteria for
  existence, stability, and robustness of a limit cycle,'' \emph{Syst. Contr.
  Lett.}, vol.~63, pp. 32--38, 2014.

\bibitem{forni2013dissip}
F.~Forni and R.~Sepulchre, ``On differentially dissipative dynamical systems,''
  \emph{{IFAC} Proceedings Volumes}, vol.~46, no.~23, pp. 15--20, 2013.

\bibitem{forni2014finslerLyap}
------, ``A differential {L}yapunov framework for contraction analysis,''
  \emph{{IEEE} Trans. Automat. Contr.}, vol.~59, no.~3, pp. 614--628, 2014.

\bibitem{manchester2018robustCCM}
I.~R. Manchester and J.-J.~E. Slotine, ``Robust control contraction metrics: A
  convex approach to nonlinear state-feedback ${H}^\infty$ control,''
  \emph{{IEEE} Contr. Syst. Lett.}, vol.~2, no.~3, pp. 333--338, 2018.

\bibitem{manchester2017control}
------, ``Control contraction metrics: Convex and intrinsic criteria for
  nonlinear feedback design,'' \emph{IEEE Trans.\ Automat.\ Contr.}, vol.~62,
  pp. 3046--3053, 2017.

\bibitem{koelewijn2021control}
P.~J. Koelewijn, R.~Toth, and S.~Weiland, ``Incremental dissipativity based
  control of discrete-time nonlinear systems via the {LPV} framework,'' in
  \emph{Proc. 60th {IEEE} Conf. Decis. Contr. ({CDC})}, 2021, pp. 3281--3286.

\bibitem{sanfelice2012metric}
R.~G. {Sanfelice} and L.~{Praly}, ``Convergence of nonlinear observers on
  $\mathbb{R}^{n}$ with a {Riemannian} metric ({Part I}),'' \emph{{IEEE}
  Trans.\ Automat.\ Contr.}, vol.~57, no.~7, pp. 1709--1722, 2012.

\bibitem{sanfelice2015convergence}
R.~G. Sanfelice and L.~Praly, ``Convergence of nonlinear observers on
  $\mathbb{R}^n$ with a {Riemannian} metric ({Part II}),'' \emph{{IEEE} Trans.\
  Automat.\ Contr.}, vol.~61, no.~10, pp. 2848--2860, 2015.

\bibitem{nijmeijer1990nonlinear}
H.~Nijmeijer and A.~van~der Schaft, \emph{Nonlinear Dynamical Control
  Systems}.\hskip 1em plus 0.5em minus 0.4em\relax Springer New York, 1990.

\bibitem{koelewijn2021incremental}
P.~J. Koelewijn and R.~T{\'{o}}th, ``Incremental stability and performance
  analysis of discrete-time nonlinear systems using the {LPV} framework,''
  \emph{{IFAC}-{PapersOnLine}}, vol.~54, no.~8, pp. 75--82, 2021.

\bibitem{parrilo2003semidefinite}
P.~A. Parrilo, ``Semidefinite programming relaxations for semialgebraic
  problems,'' \emph{Math. Program.}, vol.~96, pp. 293--320, 2003.

\bibitem{wei2021contraction}
L.~Wei, R.~McCloy, and J.~Bao, ``Contraction analysis and control synthesis for
  discrete-time nonlinear processes,'' \emph{J. Proc. Contr.}, vol. 115, pp.
  58--66, 2022.

\bibitem{verhoek2020convex}
C.~Verhoek, P.~J.~W. Koelewijn, S.~Haesaert, and R.~Toth, ``Convex incremental
  dissipativity analysis of nonlinear systems,'' \emph{Automatica}, vol. 150,
  p. 110859, 2023.

\bibitem{vilms1970totally}
J.~Vilms, ``Totally geodesic maps,'' \emph{J. Differ. Geom.}, vol.~4, no.~1,
  1970.

\bibitem{andersson2019casadi}
J.~A. Andersson, J.~Gillis, G.~Horn, J.~B. Rawlings, and M.~Diehl, ``{CasADi}:
  a software framework for nonlinear optimization and optimal control,''
  \emph{Math. Program. Comput.}, vol.~11, no.~1, pp. 1--36, 2019.

\bibitem{waechter2005ipopt}
A.~Wächter and L.~T. Biegler, ``On the implementation of an interior-point
  filter line-search algorithm for large-scale nonlinear programming,''
  \emph{Math. Program.}, vol. 106, no.~1, pp. 25--57, 2005.

\bibitem{loefberg2009sos}
J.~Löfberg, ``Pre- and post-processing sum-of-squares programs in practice,''
  \emph{{IEEE} Trans. Automat. Contr.}, vol.~54, no.~5, pp. 1007--1011, 2009.

\bibitem{mosek2019}
{MOSEK ApS}, \emph{The MOSEK optimization toolbox for MATLAB manual. Version
  9.0.}, 2019.

\bibitem{kai2017quadrotor}
J.-M. Kai, G.~Allibert, M.-D. Hua, and T.~Hamel, ``Nonlinear feedback control
  of quadrotors exploiting first-order drag effects,''
  \emph{{IFAC}-{PapersOnLine}}, vol.~50, no.~1, pp. 8189--8195, 2017.

\bibitem{nascimento2019position}
T.~P. Nascimento and M.~Saska, ``Position and attitude control of multi-rotor
  aerial vehicles: A survey,'' \emph{Annual Reviews in Control}, vol.~48, pp.
  129--146, 2019.

\bibitem{gallot2004riemannian}
S.~Gallot, D.~Hulin, and J.~Lafontaine, \emph{Riemannian Geometry}.\hskip 1em
  plus 0.5em minus 0.4em\relax Springer Berlin Heidelberg, 2004.

\end{thebibliography}
\end{document}